\documentclass[aps,prd,floatfix,nofootinbib,showpacs,twocolumn,superscriptaddress]{revtex4-1}

\usepackage{mathrsfs}
\usepackage{amsmath}
\usepackage{amssymb}
\usepackage{bbold}
\usepackage{color}
\usepackage{graphicx}
\usepackage{soul}

\usepackage[mathscr,scaled=1.15]{urwchancal}
\DeclareFontFamily{OT1}{pzc}{}
\DeclareFontShape{OT1}{pzc}{m}{it}%
{<-> s * [1.15] pzcmi7t}{}
\DeclareMathAlphabet{\mathpzc}{OT1}{pzc}{m}{it}

\definecolor{purple}{rgb}{0.5,0,0.5}
\definecolor{blue}{rgb}{0.0,0,0.9}

\begin{document}

\title{Kaon and pion parton distribution amplitudes to twist-three}

\author{Chao Shi}
\affiliation{Key Laboratory of Modern Acoustics, MOE, Institute of Acoustics, Nanjing University, Nanjing 210093, China}
\affiliation{Department of Physics, Nanjing University, Nanjing 210093, China}

\author{Chen Chen}
\affiliation{Hefei National Laboratory for Physical Sciences at the Microscale,
University of Science and Technology of China, Hefei, Anhui 230026, P. R. China}
\affiliation{Institute for Theoretical Physics and Department of Modern Physics,\\
University of Science and Technology of China, Hefei, Anhui 230026, P. R. China}

\author{Lei Chang}
\affiliation{CSSM, School of Chemistry and Physics
University of Adelaide, Adelaide SA 5005, Australia}

\author{Craig D.\ Roberts}
\affiliation{Physics Division, Argonne National Laboratory, Argonne, Illinois 60439, USA}

\author{Sebastian M.\ Schmidt}
\affiliation{Institute for Advanced Simulation, Forschungszentrum J\"ulich and JARA, D-52425 J\"ulich, Germany}

\author{Hong-Shi Zong}
\affiliation{Department of Physics, Nanjing University, Nanjing 210093, China}

\date{30 March 2015}

\begin{abstract}
$\,$\\[-5ex]\hspace*{\fill}{\emph{Preprint no}. ADP-15-12/T914}\\[1ex]
We compute all kaon and pion parton distribution amplitudes (PDAs) to twist-three and find that only the pseudotensor PDA can reasonably be approximated by its conformal limit expression.  At terrestrially accessible energy scales, the twist-two and pseudoscalar twist-three PDAs differ significantly from those functions commonly associated with their forms in QCD's conformal limit.  In all amplitudes studied, $SU(3)$ flavour-symmetry breaking is typically a 13\% effect.  This scale is determined by nonperturbative dynamics; namely, the current-quark-mass dependence of dynamical chiral symmetry breaking.  The heavier-quark is favoured by this distortion, for example, support is shifted to the $s$-quark in the negative kaon.  It appears, therefore, that at energy scales accessible with existing and foreseeable facilities, one may obtain reliable expectations for experimental outcomes by using these ``strongly dressed'' PDAs in formulae for hard exclusive processes.  Following this procedure, any discrepancies between experiment and theory will be significantly smaller than those produced by using the conformal-limit PDAs.  Moreover, the magnitude of any disagreement will either be a better estimate of higher-order, higher-twist effects or provide more realistic constraints on the Standard Model.
\end{abstract}

\pacs{
14.40.Be,   
14.40.Df,   
12.15.Hh,   
12.38.Lg    
}

\maketitle

\section{Introduction}
Knowing the behaviour of kaon and pion light-front parton distribution amplitudes (PDAs) is crucial in the analysis of a wide variety of hard exclusive processes.  For instance, the twist-two amplitudes modulate the evolution of kaon and pion elastic electromagnetic form factors \cite{Lepage:1979zb,Efremov:1979qk,Lepage:1980fj}; and both twist-two and -three amplitudes are critical in the study of nonleptonic $B$-meson decays and their use in constraining elements of the CKM matrix \cite{Beneke:2001ev}.  In analyses of this type, owing to the absence of reliable computations of the PDAs, it has been common to employ PDAs appropriate to the conformal limit of QCD when estimating the size of nonperturbative factors appearing in the factorisation formula for a given exclusive process.  For example, the normalisation expected for the asymptotic behaviour of the pion's elastic electromagnetic form factor, $F_\pi(Q^2)$, has typically been based on the following conformal form for the twist-two PDA  \cite{Brodsky:1980ny,Braun:2003rp}: $\varphi^{\rm cl}_\pi(u)= 6 u(1-u)$.  This approach to the estimation of soft factors in hard-process factorisation formulae may now be reconsidered because it has become possible both to calculate the pointwise behaviour of meson PDAs using continuum methods in QCD \cite{Chang:2013pq,Chang:2013epa,Gao:2014bca} and to validate those results via comparisons with analyses \cite{Cloet:2013tta,Segovia:2013eca} of low moments of these PDAs computed using lattice-QCD (lQCD).

An illustration of the impact of these advances is provided by a recent study of $F_\pi(Q^2)$ \cite{Chang:2013nia}.  Specifically, it has been shown \cite{Chang:2013pq} that, at all momentum scales $Q^2\gg \Lambda_{\rm QCD}^2$ which are achievable using existing or planned facilities, the conformal-limit twist-two PDA, $\varphi^{\rm cl}_\pi(x)$, provides a poor approximation to the light-meson twist-two amplitudes.  Instead, the PDA is a broad, concave function, \emph{viz}.\ $\varphi_\pi(x) \sim (8/\pi) \sqrt{x (1 - x)}$.  Evidence in support of this character had long been accumulating \cite{Mikhailov:1986be,Petrov:1998kg,Braun:2006dg,Brodsky:2006uqa} but overlooked.  Using such a dilated PDA, the soft normalisation factor associated with the asymptotic behaviour of $F_\pi(Q^2)$ increases by a factor $2$-$3$.  Thus a mismatch, which had long appeared to be a serious discrepancy between contemporary data and direct calculations on one hand, and the result obtained via the factorisation formula on the other, is transformed into near agreement.  Hence, in what may be a significant boost to the programme \cite{Dudek:2012vr,Brodsky:2015aia}, experiments at the upgraded Jefferson Laboratory (JLab) \cite{E1206101,E1207105} will potentially see a clear sign of parton model scaling for the first time in a hadron elastic form factor.

It is worth highlighting that the dilation of $\varphi_\pi(x)$ is a clear expression of dynamical chiral symmetry breaking (DCSB) on the light-front.  Consequently, empirical verification at JLab of the predicted normalisation of $F_\pi(Q^2)$ \cite{Chang:2013nia} will provide novel insights into the mechanism that may be identified with the generation of more than 98\% of the proton's mass.  In a related vein, it has been argued \cite{Chang:2013epa} that one of the two twist-three pion distribution amplitudes, \emph{i.e}.\ the pseudoscalar projection of the pion's light-front wave function, $\omega_\pi(x)$, may be understood to describe the probability distribution of the chiral condensate within the pion \cite{Maris:1997tm,Brodsky:2009zd,Brodsky:2010xf,Chang:2011mu,Brodsky:2012ku}.  Given the pseudo-Goldstone-boson character of the pion and kaon, a comparison between the pointwise behaviour of this distribution amplitude within these mesons is particularly interesting and potentially instructive.

The behaviour of the kaon's twist-two PDA, $\varphi_K(x)$, is also determined primarily by DCSB \cite{Segovia:2013eca,Shi:2014uwa}.  Indeed, $\varphi_K(x)$ is a broad, concave and asymmetric function, whose peak is shifted 12-16\% away from its position in QCD's conformal limit.  These features show that the heavier quark in the kaon carries more of the bound-state's momentum than the lighter quark; and also that DCSB modulates the magnitude of flavour-symmetry breaking because it is markedly smaller than one might expect based on the difference between light-quark current masses.  Combining these features of $\varphi_K(x)$ with those of $\varphi_\pi(x)$ described above, one obtains an improved understanding of the ratio of kaon and pion electromagnetic form factors measured at large timelike momenta \cite{Seth:2012nn}: it eliminates much of the discrepancy between experiment and theory which appears if the conformal-limit kaon and pion PDAs are used in the relevant hard scattering formulae \cite{Shi:2014uwa}.

It would thus appear that a consistent picture is emerging from the confluence between continuum and lattice QCD studies regarding the character of light-meson twist-two PDAs.  Namely, that at energy scales accessible with existing and foreseeable facilities, reliable insights concerning the Standard Model may only be obtained by using the broad, concave PDAs whose nature and origin we have indicated above.  Herein, therefore, we present calculations and results for all six kaon and pion two-particle distribution amplitudes that appear to twist-three in an expansion of the light-front wave functions of these pseudoscalar mesons.  These amplitudes have previously been estimated using their properties under conformal transformations as the guiding principle in concert with QCD sum rules \cite{Braun:1989iv,Ball:2006wn}; but our analysis is the first to use the Dyson-Schwinger equations (DSEs) \cite{Maris:2003vk,Chang:2011vu,Bashir:2012fs,Cloet:2013jya}, which have both a direct connection with QCD and can completely chart the pointwise behaviour of these amplitudes.

This document is organised as follows.  In Sec.\,\ref{SectionTwo} we introduce the distribution amplitudes and describe the method that will be used in their computation.  Section\,\ref{secResults} provides an algebraic illustration of our techniques.  The algebraic formulae we obtain also serve as a benchmark against which to evaluate the nature of our numerical results, described in detail in Sec.\,\ref{resultsnumerical}.  A summary and perspective are presented in Sec.\,\ref{epilogue}.

\section{Distribution Amplitudes and Bethe-Salpeter Wave Functions}
\label{SectionTwo}
\subsection{Definitions and Observations}
A pseudoscalar meson, $P_{\bar g f}(q)$, with mass $m_P$, possesses three two-particle light-cone distribution amplitudes to twist-three, which may be expressed thus \cite{Braun:1989iv}:
\begin{subequations}
\label{PSPDAs}
\begin{align}
\nonumber
\lefteqn{\langle 0|\bar{\psi}_f(-x)\gamma_5\gamma\cdot n \psi_{g}(x)|P_{\bar g f}(q)\rangle}\\
&= f_P \, n\cdot q \int^1_0 du \, {\rm e}^{-i x\cdot q\, (2 u-1)}
\varphi_P^{(2)}(u,\zeta)\,, \label{phi2}\\
\nonumber
-\lefteqn{\langle 0|\bar{\psi}_f(-x)i\gamma_5 \psi_{g}(x)|P_{\bar g f}(q)\rangle} \\
&= i \rho_P^\zeta \int^1_0 du \, {\rm e}^{-i x\cdot q\, (2 u-1)} \omega_P^{(3)}(u,\zeta)\,, \label{T35}\\
\nonumber
\lefteqn{\langle 0|\bar{\psi}_f(-x)i\gamma_5\sigma_{\mu\nu}  q_\mu n_\nu \psi_{g}(x)|P_{\bar g f}(q)\rangle}\\
 &= \frac{1}{4} \rho_P^\zeta \,n\cdot q \int^1_0 du \, {\rm e}^{-i x\cdot q\, (2 u-1)} \frac{d}{du}\upsilon_P^{(3)}(u,\zeta)\,.\label{T3T}
\end{align}
\end{subequations}
Here $q^2=-m_P^2$;\footnote{We use a Euclidean metric:  $\{\gamma_\mu,\gamma_\nu\} = 2\delta_{\mu\nu}$; $\gamma_\mu^\dagger = \gamma_\mu$; $\gamma_5= \gamma_4\gamma_1\gamma_2\gamma_3$, tr$[\gamma_5\gamma_\mu\gamma_\nu\gamma_\rho\gamma_\sigma]=-4 \epsilon_{\mu\nu\rho\sigma}$; $\sigma_{\mu\nu}=(i/2)[\gamma_\mu,\gamma_\nu]$; $a \cdot b = \sum_{i=1}^4 a_i b_i$; and $q_\mu$ timelike $\Rightarrow$ $q^2<0$.}
$x_\mu = (z/2) n_\mu$, with $n$ a light-like four-vector, $n^2=0$, $n\cdot q = -m_P$;
the superscript labels the twist-order, which will be omitted hereafter;
$\zeta$ is the renormalisation scale;
and $f_P$, $\rho_P^\zeta$ are, respectively, the pseudovector and pseudoscalar projections of the meson's Bethe-Salpeter wave function onto the origin in configuration space, explicit expressions for which are given in Eqs.\,\eqref{normconstants} below.  With the conventions specified by Eqs.\,\eqref{PSPDAs}, each of the PDAs is unit normalised, \emph{viz}.\
\begin{equation}
\label{normalised}
\int_0^1 du \, \{ \varphi_P(u;\zeta)\,,\,\omega_P(u;\zeta)\,,\, \upsilon_P(u;\zeta)\} = 1\,.
\end{equation}
The reason why we have expressed the left-hand-side of Eq.\,\eqref{T3T} in terms of a differentiated PDA will subsequently become apparent -- see, \emph{e.g}., Eq.\,\eqref{upsilonasy} and the preceding analysis.  In considering the twist-three amplitudes, Eqs.\,\eqref{T35} and \eqref{T3T}, one should bear in mind that they are not truly independent: a three-particle (quark$+$antiquark$+$gluon) twist-three amplitude connects them \cite{Braun:1989iv,Braun:2003rp}.

Note that in order to produce quantities that are gauge invariant for all values of $x$, each of the left-hand-sides in Eqs.\,\eqref{PSPDAs} should also contain a Wilson line:
\begin{equation}
{\cal W}[-x,x] = \exp ig\int_{-x}^{x} d \sigma_\mu A_\mu(\sigma)\,,
\end{equation}
between the quark fields.  Plainly, for any light-front trajectory, ${\cal W}[-x,x]\equiv 1$ in lightcone gauge: $n\cdot A=0$, and hence the Wilson line does not contribute when this choice is employed.  On the other hand, light-cone gauge is seldom practicable in either model calculations or quantitative nonperturbative analyses in continuum QCD.  In fact, herein, as is typical in nonperturbative DSE studies, we employ Landau gauge because, \emph{inter alia} \cite{Bashir:2008fk,Bashir:2009fv,Raya:2013ina}: it is a fixed point of the renormalisation group; that gauge for which sensitivity to model-dependent differences between \emph{Ans\"atze} for the fermion--gauge-boson vertex are least noticeable; and a covariant gauge, which is readily implemented in numerical simulations of lattice-regularised QCD.

We therefore proceed by assuming that ${\cal W}[-x,x]$ is not quantitatively important in computation of the two-particle amplitudes in Eqs.\,\eqref{PSPDAs}.  That has been verified for $\varphi_P$ in Eq.\,\eqref{phi2} \cite{Kopeliovich:2011rv} and it is plausible for $\omega_P$, $\upsilon_P$.  However, this omission should be borne in mind; and the validity of the assumption judged through comparisons with results obtained using other methods which are also soundly grounded in QCD.

The value of $\zeta$ in Eqs.\,\eqref{PSPDAs} specifies the mass-scale relevant to the process in which the meson is involved and hence at which the PDA is to be employed.  The shape of a given PDA changes with $\zeta$; hence the $\zeta$-dependence of the PDAs is important.  The evolution equation for two-particle twist-two distributions is known in closed form and it has the solution \cite{Lepage:1979zb,Efremov:1979qk,Lepage:1980fj}:
\begin{align}
\label{PDAG3on2}
\varphi_P(u;\tau) & = \varphi^{\rm cl}(u)
\sum_{j=0,1,2,\ldots}^{\infty} \!\! \!\! a_j^{3/2}(\tau) \,C_j^{(3/2)}(u -\bar u) ,\;\;\\
\varphi^{\rm cl}(u) & = 6 u (1-u)\,, \label{phiasy}
\end{align}
where $\tau=1/\zeta$, $\bar u= 1-u$, $a_0=1$.  The expansion coefficients $\{a_j^{3/2},j\geq 1\}$ evolve logarithmically with $\tau$: they vanish as $\tau\to 0$.  These features owe to the fact that, on $\tau \Lambda_{\rm QCD} \simeq 0$, QCD is invariant under the collinear conformal group
SL$(2;\mathbb{R})$
\cite{Brodsky:1980ny,Braun:2003rp}.  The Gegenbauer-$\alpha=3/2$ polynomials correspond to irreducible representations of this group and hence the expansion in Eq.\,\eqref{PDAG3on2}.  

The evolution of the two-particle twist-three distributions is significantly more complicated because they are related to the matrix elements of three-particle (quark$+$antiquark$+$gluon) distributions, which have nontrivial scale dependence and mix with each other under renormalisation \cite{Braun:1989iv,Ball:1998je,Braun:2003rp,Ball:2006wn,Kim:2008ir}.  We therefore omit further discussion of evolution and focus instead on reporting result computed with
\begin{equation}
\zeta=\zeta_2 :=2\,\,{\rm GeV},
\end{equation}
a value chosen because it is typically that employed in numerical simulations of lQCD.

It is worth remarking that until recently it was commonly assumed that at any length-scale $\tau=1/\zeta$, an accurate approximation to a given PDA was obtained using just the first few terms in the associated conformal expansion.  In connection with the twist-two distribution in Eq.\,\eqref{phi2}, this amounted to the assumption that an accurate representation of $\varphi_P(x;\tau)$ was obtained by using just the first few terms of the expansion in Eq.\,\eqref{PDAG3on2}.  Let us call this \emph{Assumption~A}.  It has led to models for $\varphi_P(x;\tau)$ whose pointwise behaviour is not concave on $x\in[0,1]$, \emph{e.g}.\ to ``humped'' distributions \cite{Chernyak:1983ej}.  Following Ref.\,\cite{Chang:2013pq}, one can readily establish that a double-humped form for $\varphi_P(x)$ lies within the class of distributions produced by a meson Bethe-Salpeter amplitude which may be characterised as vanishing at zero relative momentum, instead of peaking thereat.  No ground-state pseudoscalar or vector meson Bethe-Salpeter equation solution exhibits corresponding behaviour \cite{Maris:1997tm,Maris:1999nt,Qin:2011xq}.

\emph{Assumption~A} is certainly valid on $\tau \Lambda_{\rm QCD} \simeq 0$.  However, it is unsound at any energy scale accessible in contemporary or foreseeable experiments.  This was highlighted in Ref.\,\cite{Cloet:2013tta} and in Sec.\,5.3 of Ref.\,\cite{Cloet:2013jya}.  The latter used the fact \cite{Georgi:1951sr,Gross:1974cs,Politzer:1974fr} that $\varphi^{\rm cl}(x)$ can only be a good approximation to a meson's PDA when it is accurate to write $u_{\rm v}(x) \approx \delta(x)$, where $u_{\rm v}(x)$ is the meson's valence-quark PDF, and showed that this is not valid even at energy scales characteristic of the large hadron collider.  An identical conclusion was reached in Refs.\,\cite{Segovia:2013eca,Shi:2014uwa}, via consideration of the first moment of the kaon's PDA, which is a direct measure of $SU(3)$ flavour-symmetry breaking and must therefore vanish in the conformal limit.  Hence, realistic two-particle twist-two meson PDAs are necessarily broader than $\varphi^{\rm cl}(x)$; in fact, much broader.  It follows that insistence on using just a few terms in Eq.\,\eqref{PDAG3on2} to represent a hadron's twist-two PDA will typically lead to unphysical oscillations, \emph{i.e}.\ humps, just as any attempt to represent a box-like curve via a Fourier series will inevitably lead to slow convergence and spurious oscillations.

The preceding observations lead one to appreciate that two-particle twist-three distributions evaluated at $\zeta_2$ need not closely resemble the functional forms associated with their conformal limits.  In our analysis of the PDAs in Eqs.\,\eqref{PSPDAs}, we will not prejudice our analysis in this way.  We will instead use an alternative to \emph{Assumption~A}, which is explained in the next subsection.

\subsection{Computing Light-front Projections of Bethe-Salpeter Wave Functions}
Returning now to Eqs.\,\eqref{PSPDAs}, accounting for Euclidean metric and making use of the relationship between Bethe-Salpeter wave functions in configuration and momentum space \cite{LlewellynSmith:1969az}, one obtains
\begin{subequations}
\label{PSPDAsE}
\begin{align}
\label{phi2E}
f_P \,\varphi_P(u)  & =
 N_c {\rm tr}\,
Z_2 \! \int_{dk}^\Lambda \!\!\delta_n^{u}(k_\eta) \,\gamma_5\gamma\cdot n\, \chi_P^q(k_\eta,k_{\bar\eta})\,,\\
\label{omegaE}
i\rho_P^\zeta \,\omega_P(u) & =
 N_c {\rm tr}\,
Z_4 \! \int_{dk}^\Lambda \!\!\delta_n^{u}(k_\eta) \,\gamma_5\, \chi_P^q(k_\eta,k_{\bar\eta})\,,\\
\frac{1}{4} \rho_P^\zeta \upsilon^\prime(u) & =
N_c {\rm tr}\,
Z_4 \! \int_{dk}^\Lambda \!\!\delta_n^{u}(k_\eta) \,\gamma_5\sigma_{\mu\nu}q_\mu n_\nu\, \chi_P^q(k_\eta,k_{\bar\eta})\,,\label{upsilonE}
\end{align}
\end{subequations}
where $N_c=3$; the trace is over spinor indices;
$\int_{dk}^\Lambda$ is a Poincar\'e-invariant regularisation of the four-dimensional integral, with $\Lambda$ the ultraviolet regularisation mass-scale;
$Z_{2,4}(\zeta,\Lambda)$ are, respectively, the quark wave-function and Lagrangian mass renormalisation constants, computed using a mass-independent renormalisation scheme \cite{Weinberg:1951ss}; and
$\delta_n^{u}(k_\eta):= \delta(n\cdot k_\eta - u \,n\cdot q)$.

The Bethe-Salpeter wave function in Eqs.\,\eqref{PSPDAsE} is
\begin{equation}
\chi_P^q(k_\eta,k_{\bar\eta}) = S_f(k_\eta) \Gamma_P(k_{\eta\bar\eta};q) S_g(k_{\bar \eta})\,,
\label{chiP}
\end{equation}
where $\Gamma_P$ is the Bethe-Salpeter amplitude; $S_{f,g}$ are the dressed-quark propagators, which are usually written in one of the following, equivalent forms
\begin{subequations}
\label{SgeneralN}
\begin{align}
S_{f}(k) &= -i \gamma\cdot k\,\sigma_V^f(k^2) + \sigma_S^f(k^2)\\
&= 1/[ i \gamma\cdot k\,A_f(k^2) + B_f(k^2)]\\
&= Z_f(k^2)/[i \gamma\cdot k + M_f(k^2)]\,, \label{MassFun}
\end{align}
\end{subequations}
and $k_{\eta\bar\eta} = [k_\eta+k_{\bar\eta}]/2$, $k_\eta = k + \eta q$, $k_{\bar\eta} = k - (1-\eta) q$, $\eta\in [0,1]$.  Owing to Poincar\'e covariance, no observable can legitimately depend on $\eta$, \emph{i.e}.\ the definition of the relative momentum.

The pseudoscalar meson Bethe-Salpeter amplitude in Eq.\,\eqref{chiP} has the form $(\ell = k_{\eta\bar\eta})$
\begin{eqnarray}
\nonumber
\lefteqn{\Gamma_{P}(\ell;q) = \gamma_5
\big[ i E_{P}(\ell;q) + \gamma\cdot P F_{P}(\ell;q)   }\\
&&  \quad\quad  + \gamma\cdot \ell \, G_{P}(\ell;q) + \sigma_{\mu\nu} \ell_\mu q_\nu H_{P}(\ell;q) \big]. 
\label{BSK}
\end{eqnarray}
Each of the scalar functions in Eq.\,\eqref{BSK} has the following decomposition
\begin{eqnarray}
\mathcal{F}(\ell;q)=\mathcal{F}_0(\ell;q)+\ell \cdot q \,\mathcal{F}_1(\ell;q)\,,
\label{decom}
\end{eqnarray}
with $\mathcal{F}_{0,1}$ even under $(\ell\cdot q) \to (-\ell\cdot q)$.  Herein, we consider the isospin-symmetric limit $m_u=m_d\neq m_s$.  Note that for mesons constituted from valence-quarks with equal current-mass, $\mathcal{F}_{1}\equiv 0$.

At this point it is useful to expose the meaning of the normalisation factors in Eqs.\,\eqref{PSPDAsE}.  To that end, we act with $\int_0^1 du$ on both sides of Eqs.\,\eqref{phi2E}, \eqref{omegaE} and employ Eqs.\,\eqref{normalised}, thereby arriving at:
\begin{subequations}
\label{normconstants}
\begin{align}
f_P & =
 N_c {\rm tr}\,
Z_2 \! \int_{dk}^\Lambda\gamma_5\gamma\cdot n\, \chi_P^q(k_\eta,k_{\bar\eta})\,,\\
i \rho_P^\zeta & =
 N_c {\rm tr}\,
Z_4 \! \int_{dk}^\Lambda \gamma_5\, \chi_P^q(k_\eta,k_{\bar\eta})\,. \label{normrhoP}
\end{align}
\end{subequations}
These expressions will readily be recognised as distinct projections onto the origin in configuration space of the meson's Bethe-Salpeter wave function, \emph{i.e}.\ the meson's pseudovector and pseudoscalar decay constants \cite{Maris:1997hd}.  Both express an intrinsic property of the meson and are equivalent order parameters for DCSB \cite{Roberts:2000aa}.  In this connection, the latter ($\rho_P^\zeta$) has been identified as the in-meson chiral condensate \cite{Maris:1997tm,Brodsky:2009zd,Brodsky:2010xf,Chang:2011mu,Brodsky:2012ku}.

As reviewed elsewhere \cite{Maris:2003vk,Chang:2011vu,Bashir:2012fs,Cloet:2013jya}, it is now possible to obtain realistic meson Bethe-Salpeter amplitudes by solving a coupled system of integral equations; namely, symmetry-preserving truncations of QCD's gap and Bethe-Salpeter equations.  That given, then, with $\chi_P$ in hand, it is straightforward to follow the procedure explained in Refs.\,\cite{Chang:2013pq,Chang:2013epa,Gao:2014bca,Shi:2014uwa} and thereby obtain the meson PDAs from Eqs.\,\eqref{PSPDAsE}.  The first step is to compute the moments
\begin{equation}
\langle u_\Delta^m \rangle_\phi  = \int_0^1 du\, (2 u-1)^m \phi_P(u)
\end{equation}
where $\phi=\varphi_P$, $\omega_P$, $\upsilon_P^\prime$, which are determined explicitly via
\begin{subequations}
\label{momentsE}
\begin{align}
\nonumber
& \rule{-0.5em}{0ex} f_P  \langle u_\Delta^m \rangle_\varphi  \\
& \rule{-1em}{0ex} = N_c {\rm tr} Z_2\int_{dk}^\Lambda {\mathpzc D}(n,k_\eta,q,m)
\,\gamma_5\gamma\cdot n\, \chi_P^q(k_\eta,k_{\bar\eta})\,,\label{phi2EE}\\
\nonumber & \rule{-0.5em}{0ex} i\rho_P^\zeta
\langle u_\Delta^m \rangle_\omega  \\
& \rule{-1em}{0ex} =  N_c {\rm tr}\,
Z_4 \! \int_{dk}^\Lambda  {\mathpzc D}(n,k_\eta,q,m) \,\gamma_5\, \chi_P^q(k_\eta,k_{\bar\eta})\,, \label{omegaEE}\\
\nonumber
& \rule{-0.5em}{0ex} \frac{1}{4} \rho_P^\zeta
\langle u_\Delta^m \rangle_{\upsilon^\prime} \\
& \rule{-1em}{0ex} =  N_c {\rm tr}\,
Z_4 \! \int_{dk}^\Lambda
{\mathpzc D}(n,k_\eta,q,m)\,\gamma_5\sigma_{\mu\nu}q_\mu n_\nu\, \chi_P^q(k_\eta,k_{\bar\eta}) \label{upsilonEE}\,,
\end{align}
\end{subequations}
where $[n\cdot q]^{m+1 }{\mathpzc D}(n,k,q,m) = [2 n\cdot k - n\cdot q]^m$.  Notably, beginning with an accurate form of $\chi_P^q$, arbitrarily many moments can be computed.\footnote{Recall that our approach is formulated in Euclidean space.  Thus, the moments provide a practical way to make the connection with Minkowski space, wherein the light-front is defined.  In cases where the propagators and vertices take a simple form, direct calculation of the PDAs is straightforward.  However, with sophisticated DSE-generated propagators and vertices, it is advantageous to employ reconstruction from the moments.}

It is now useful to write
\begin{subequations}
\label{EOform}
\begin{eqnarray}
\phi_P(u) &= &  \phi_P^E(u) + \phi_P^O(u)\,,\\
\phi_P^{E,O}(u) & = & (1/2)[\phi_P(u)\pm \phi_P(\bar u)] \,, 
\end{eqnarray}
\end{subequations}
in which form the nonzero moments of $\phi_P^E(u)$ reproduce all the $m$-even moments of $\phi_P$ and the nonzero moments of $\phi_P^O(u)$ are the $m$-odd moments of $\phi_P$.  Plainly, $\phi_P^O(u)\equiv 0$ for mesons comprised from valence-quarks with degenerate current-masses; but Eqs.\,\eqref{EOform} enable us to describe the treatment of all systems simultaneously.

Consider now that Gegenbauer polynomials of order $\alpha$, $\{C_n^{\alpha}(2 u -1)\,|\, n=0,\ldots,\infty\}$, are a complete orthonormal set on $u\in[0,1]$ with respect to the measure $[u (1-u)]^{\alpha_-}$, $\alpha_-=\alpha-1/2$.  They therefore enable reconstruction of any function defined on $u\in[0,1]$; and hence, with complete generality and to a level of accuracy defined by the summation upper bounds,
\begin{equation}
\label{PDAGalpha}
\phi_P^{E,O}(u) \approx \,_m\phi_P^{E,O}(u) \,,
\end{equation}
where
\begin{subequations}
\label{phimEO}
\begin{eqnarray}
\,_m\phi_P^E(u) &= &
 N_{\bar \alpha} \, [u (1-u)]^{\bar\alpha_-}\!\!\!\!\!
\sum_{j=0,2,4,\ldots}^{\bar j_{\rm max}} a_j^{\bar\alpha} C_j^{\bar\alpha}(2 u -1)
\,, \quad\quad \\
\,_m\phi_P^O(u) &=& N_{\hat \alpha} \, [u (1-u)]^{\hat \alpha_-}\,
\sum_{j=1,3,\ldots}^{\hat{j}_{\rm max}+1} a_j^{\hat\alpha} C_j^{\hat\alpha}(2 u -1)\,,\quad\quad
\end{eqnarray}
\end{subequations}
$N_\alpha = \Gamma(2\alpha+1)/[\Gamma(\alpha+1/2)]^2$ and $a_0^{\bar\alpha} = 1$.  In general, $\bar\alpha \neq \hat\alpha$ because $\phi_P^E(u)$ and $\phi_P^O(u)$ are orthogonal components of $\phi_P(u)$.

At this point, from a given set of $2 m_{\rm max}$ moments computed via Eqs.\,\eqref{momentsE}, the even and odd component-PDAs may be determined independently by separately minimising
\begin{subequations}
\begin{eqnarray}
\varepsilon_m^{E} &=& \sum_{l=2,4,\ldots, 2 m_{\rm max}} |\langle  u_\Delta^l\rangle_{m}^E/\langle u_\Delta^l\rangle_\phi-1|\,,\\
\varepsilon_m^O &=& \sum_{l=1,3,\ldots, 2 m_{\rm max}-1} |\langle  u_\Delta^l\rangle_{m}^O/\langle u_\Delta^l\rangle_\phi-1|\,,
\end{eqnarray}
\end{subequations}
over $\{ \bar\alpha, a_2, a_4, \ldots, a_{j_{\rm max}}\}$, $\{ \hat \alpha, a_1, a_3, \ldots, a_{j_{\rm max}+1}\}$,  where
\begin{equation}
\langle u_\Delta^l\rangle_{m}^{E,O} = \int_0^1 du\, (2u-1)^l \,_m\phi_P^{E,O}(u)\,.
\label{endprocedure}
\end{equation}

This is the alternative to \emph{Assumption A} mentioned above and exploited elsewhere \cite{Chang:2013pq,Cloet:2013tta,Chang:2013epa,Cloet:2013jya,%
Chang:2013nia,Segovia:2013eca,Gao:2014bca}.  It acknowledges that at all empirically accessible scales the pointwise profile of PDAs is determined by nonperturbative dynamics; and hence PDAs should be reconstructed from moments by using Gegenbauer polynomials of order $\alpha$, with this order -- the value of $\alpha$ -- determined by the moments themselves, not fixed beforehand.  In known cases, involving $\pi$-, $K$-, $\rho$- and $\phi$-mesons, this procedure converges rapidly: $j_{\rm max}=2$ is sufficient \cite{Chang:2013pq,Chang:2013epa,Gao:2014bca,Shi:2014uwa}.

\section{Results: Algebraic Benchmarks}
\label{secResults}
%
In order to reliably compute moments via Eqs.\,\eqref{momentsE}, we follow the approach introduced in Ref.\,\cite{Chang:2013pq} and develop a Nakanishi-like representation \cite{Nakanishi:1963zz,Nakanishi:1969ph,Nakanishi:1971} of the meson Bethe-Salpeter wave functions that are ultimately obtained via numerical solution of a Bethe-Salpeter equation.  Before detailing the results of such numerical analysis, we judge it useful to provide an algebraic illustration of this idea.

Consider, therefore,  Eq.\,\eqref{upsilonEE}, treat $f=g$ and introduce
\begin{subequations}
\label{NakanishiASY}
\begin{align}
\label{eq:sim1}
S(k)&=[-i\gamma\cdot k+M]\Delta_M(k^2)\,,\\
\label{eq:sim3}
\varsigma_\nu(z)&=\frac{1}{\sqrt{\pi}}\frac{\Gamma(\nu+3/2)}{\Gamma(\nu+1)}(1-z^2)^\nu\,,\\
\label{eq:sim2}
\Gamma_P(k;q)&=i\gamma_5\frac{M}{f_P}\int^1_{-1}dz\,\varsigma_\nu(z) M^{2\nu}\hat\Delta^\nu_M(k^2_{+z})\,,
\end{align}
\end{subequations}
where $\Delta_M(s)=1/[s+M^2]$, $\hat\Delta_M(s)=M^2\Delta_M(s)$, $k_{+z}=k-(1-z)q/2$.  (Equations~\eqref{phi2EE}, \eqref{omegaEE} have been analysed in this way elsewhere \cite{Chang:2013pq,Chang:2013epa}.) Using a Feynman parametrisation in the resulting expression, the three denominators appearing in Eq.\,\eqref{upsilonEE} can be combined into one $k$-quadratic form, raised to a power that depends linearly on $\nu$.  A subsequent change of variables enables one to isolate the $d^4k$ integration and arrive at
\begin{align}
& -\frac{1}{4}\rho_P \int_0^1 du \, u^m \upsilon_P^\prime(u)
 = \frac{1}{4}\rho_P\, m \int_0^1 du \, u^{m-1} \upsilon_P(u) \\
\nonumber
&= \frac{1}{4} \bigg[4 N_c Z_4 \int_{dk}^\Lambda
     \frac{M^{1+2\nu}}{(k^2+M^2)^{2+\nu}}\bigg]\frac{\Gamma(2+\nu)}{\Gamma(\nu)}\\
&     \quad \times  \!\int_{-1}^1 \!\! dz\!\! \int_0^1 \! d\alpha d\beta \,
    \varsigma_\nu(z) \,\alpha \,(1-\alpha)^{\nu-1} \, m \,{\mathpzc A}(z,\alpha,\beta)^{m-1}\,, \label{upsilonA}
\end{align}
where ${\mathpzc A}(z,\alpha,\beta) = (1/2)(1+\alpha - 2\alpha\beta-z(1-\alpha))$.

The expression within the parentheses in Eq.\,\eqref{upsilonA} is recognisable as the value of $\rho_P$ that one obtains from Eq.\,\eqref{normrhoP} using Eqs.\,\eqref{NakanishiASY}.  It follows that
\begin{align}
\nonumber
& \int_0^1 du \, u^{m} \upsilon_P(u) = \nu (1+\nu)\!\\
\nonumber
& \times \!\int_{-1}^1 \!\! dz\!\! \int_0^1 \! d\alpha d\beta \,
    \varsigma_\nu(z) \,\alpha \,(1-\alpha)^{\nu-1} \, {\mathpzc A}(z,\alpha,\beta)^{m}\\
& = \frac{\Gamma (2 \nu +2) \Gamma (m+\nu +1)}{\Gamma (\nu +1) \Gamma (m+2 \nu +2)}\,,
\end{align}
which are the moments of
\begin{equation}
\upsilon_P^{\rm asy}(u;\nu) =  \frac{\Gamma(2+\nu)}{\Gamma(1+\nu)^2} (u\bar u)^\nu\,.
\end{equation}

A QCD-like theory is obtained with $\nu=1$, in which case
\begin{equation}
\label{upsilonasy}
\upsilon_P^{\rm asy}(u;\nu=1) = 6 u \bar u = \varphi^{\rm cl}(u)\,.
\end{equation}
It is notable that similar reasoning leads to the same result for the twist-two PDA \cite{Chang:2013pq}:
\begin{equation}
\label{varphiasy}
\varphi_P^{\rm asy}(u;\nu=1) = 6 u \bar u = \varphi^{\rm cl}(u)\,.
\end{equation}
There are two statements here.  First, the algebraic model in Eqs.\,\eqref{NakanishiASY} produces an identical result for both the twist-two and pseudotensor twist-three PDAs.  Second, both this equality and the functional form of the identical results are precisely the outcomes argued to arise in the conformal limit of QCD \cite{Braun:1989iv}.

The model defined by Eqs.\,\eqref{NakanishiASY} has also been used to evaluate an asymptotic form for $\omega_P$, with the result \cite{Chang:2013epa}
\begin{eqnarray}
\nonumber
\omega_P^{\rm asy}(u;\nu) & = & \frac{(1+\nu) \Gamma(2+2\nu)}{2 (1+2\nu) \Gamma(\nu) \Gamma(2+\nu)} \, [u\bar u]^{\nu-1}\\
&& \times \bigg[1+
\frac{C_2^{(\nu-1/2)}(u-\bar u )}{(2\nu - 1 )(\nu+1)}  \bigg]\,. \label{omeganu}
\end{eqnarray}
In this case, $\nu=1$ produces
\begin{equation}
\label{otherasy}
\omega_P^{\rm asy}(u;\nu=1) = 1 + \frac{1}{2} C_2^{(1/2)}(u-\bar u)\,.
\end{equation}
Although the functions $\{C_j^{(1/2)},j=2,4,\ldots\}$ are expected in the expansion of $\omega_P^{\rm asy}(u)$ on the domain $\tau \Lambda_{\rm QCD} \simeq 0$, Eq.\,\eqref{otherasy} is not the functional form anticipated of QCD's conformal-limit, which is instead \cite{Braun:1989iv}:
\begin{equation}
\label{othercl}
\omega_P^{\rm cl}(u)\equiv 1\,.
\end{equation}
An explanation for this difference might be found in the conformal expansion of $\omega_P(u)$ provided in Eq.\,(33) of Ref.\,\cite{Braun:1989iv}.  There, the coefficient of the $C_2^{(1/2)}$ term is proportional to $f_{3\pi}$, a leading moment of the pion's three-particle (quark$+$antiquark$+$gluon) twist-three amplitude, which vanishes logarithmically as the renormalisation scale, $\zeta$, is removed to infinity.  From this perspective, the term $(1/2) C_2^{(1/2)}$ in Eq.\,\eqref{otherasy} and the absence of an analogous correction in Eq.\,\eqref{upsilonasy} are a statement that Eqs.\,\eqref{NakanishiASY} implicitly express a mixing pattern between the two- and three-particle twist-three amplitudes.

We will subsequently provide comparisons between the forms in Eqs.\,\eqref{upsilonasy}--\eqref{otherasy} and results for all the PDAs computed using realistic dressed-quark propagators and Bethe-Salpeter amplitudes.  The differences will expose some of the impacts of nonperturbative dynamics and/or violations of $SU(3)$ flavour-symmetry in QCD's twist-two and -three two-particle sectors.  Crucially, as we shall see, $\omega_P^{\rm asy}(u;\nu=1)$ in Eq.\,\eqref{otherasy} and $\upsilon_P^{\rm asy}(u;\nu=1)$ in Eq.\,\eqref{upsilonasy} are the natural benchmarks for all existing, realistic studies of two-particle, twist-three PDAs that make reference to accessible energy scales.

\section{Results: Numerical Computations}
\label{resultsnumerical}
\subsection{Quark propagators and meson Bethe-Salpeter amplitudes}
We solved the $s$- and $u$- quark gap equations and the kaon and pion Bethe-Salpeter equations numerically, using the interaction in Ref.\,\cite{Qin:2011dd}.  The infrared composition of this interaction is deliberately consistent with that determined in modern studies of QCD's gauge sector \cite{Bowman:2004jm,Cucchieri:2011ig,Boucaud:2011ug,Ayala:2012pb,%
Aguilar:2012rz,Strauss:2012dg}; and, in the ultraviolet, it preserves the one-loop renormalisation group behaviour of QCD so that, \emph{e.g}., the dressed-quark mass-functions, $M_{s,u}(p^2)$ in Eq.\,\eqref{MassFun}, are independent of the renormalisation point, which we choose to be $\zeta_2$.

In completing the gap and Bethe-Salpeter kernels we employ two different procedures and compare their results: rainbow-ladder (RL) truncation, detailed in App.\,A.1 of Ref.\,\cite{Chang:2012cc}, which is the most widely used DSE computational scheme in hadron physics, whose strengths and weakness are canvassed elsewhere \cite{Maris:2003vk,Chang:2011vu,Bashir:2012fs,Cloet:2013jya}; and the modern DCSB-improved (DB) kernels detailed in App.\,A.2 of Ref.\,\cite{Chang:2012cc}, which are the most refined kernels currently available \cite{Chang:2009zb,Chang:2010hb,Chang:2011ei,Cloet:2013jya}.  Both schemes are symmetry-preserving; but the latter introduces essentially nonperturbative DCSB effects into the kernels, which are omitted in RL truncation and any stepwise improvement thereof.  The DB kernel is thus the more realistic.

As detailed elsewhere \cite{Binosi:2014aea}, this conclusion is supported by the agreement emerging between the ``top-down'' approach to determining the quark-quark interaction in QCD, which works toward an \emph{ab initio} computation of the interaction via direct analysis of the gauge-sector gap equations, and the DB kernel determined via the ``bottom-up'' scheme, which aims to infer the interaction by fitting data within a well-defined truncation of those equations in the matter sector that are relevant to bound-state properties.

The gap and Bethe-Salpeter equation solutions are obtained as matrix tables of numbers.  Computation of the moments in Eqs.\,\eqref{momentsE} is cumbersome with such input, so we employ algebraic parametrisations of each array to serve as interpolations in evaluating the moments.  For the quark propagators, we represent $\sigma_{V,S}$ as meromorphic functions with no poles on the real $k^2$-axis \cite{Bhagwat:2002tx}, a feature consistent with confinement as defined through the violation of reflection positivity \cite{Gribov:1999ui,Krein:1990sf,%
Dokshitzer:2004ie,Roberts:2007ji,Chang:2011vu,Bashir:2012fs,%
Cloet:2013jya}.  Concerning the Bethe-Salpeter amplitudes, each scalar function in Eq.\,\eqref{BSK} is expressed via a Nakanishi-like representation \cite{Nakanishi:1963zz,Nakanishi:1969ph,Nakanishi:1971}, with parameters fitted to that function's first two (pion) or four (kaon) $\ell\cdot q$ Chebyshev moments.  The quality of the description is illustrated via the dressed-quark propagator in Fig.\,\ref{fig:Splot}. (Details of these procedures are presented in Appendix~\ref{sec:interpolations}.)

\begin{figure}[t]

\centerline{\includegraphics[width=0.78\linewidth]{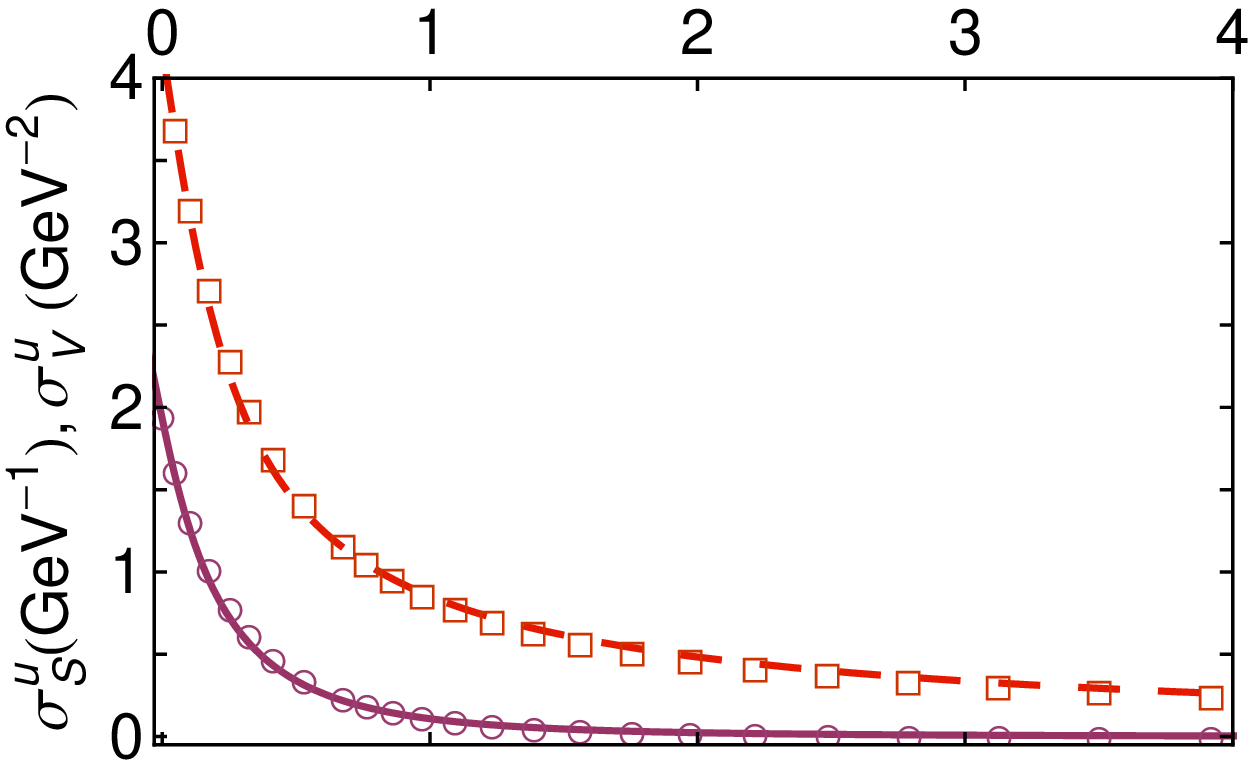}}

\vspace*{-8.3ex}
\centerline{\includegraphics[width=0.78\linewidth]{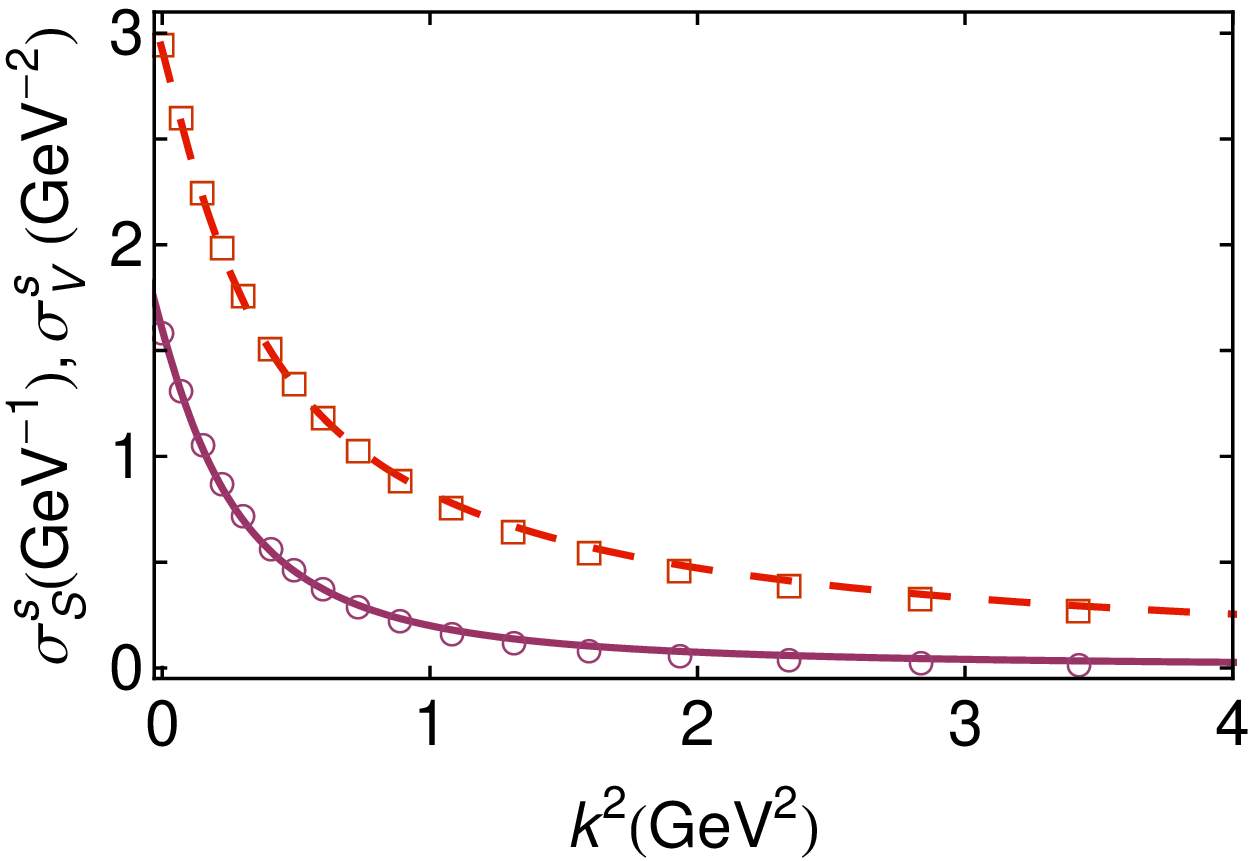}}

\caption{Functions characterising the dressed quark propagator in the DB truncation.
\emph{Upper panel}. $u/d$-quark functions, $\sigma_{S,V}^{u/d}(p^2)$ -- solution (open circles and squares, respectively) and interpolation functions (solid and long-dashed curves, respectively).
\emph{Lower panel}. $s$-quark\ functions, $\sigma_{S,V}^s(p^2)$.  Same legend.
\label{fig:Splot}}
\end{figure}

Using the interpolating spectral representations, it is straightforward to compute arbitrarily many moments of the meson PDAs via Eqs.\,\eqref{momentsE}. We typically employ $2 m_{\rm max}=50$.  The pointwise forms of the PDAs are then reconstructed via the ``Gegenbauer-$\alpha$'' procedure described in connection with Eqs.\,\eqref{PDAGalpha}--\eqref{endprocedure} above.  Again, the procedure converges rapidly in all cases, so that results obtained with $j_{\rm max}=2$ produce $\varepsilon_m^{E,O} < 1$\%.

\subsection{Twist-two}
The two-particle leading-twist PDAs for the pion and negative-kaon\footnote{The PDA for the corresponding antiparticle is obtained through the replacement $u\to \bar u=1-u$.}
were computed and discussed in Refs.\,\cite{Chang:2013pq,Segovia:2013eca,Shi:2014uwa}.  The results obtained with the DB kernel are
\begin{equation}
\label{pionPDA2}
\varphi_\pi^{\rm DB}(u;\zeta_2) = 1.81 [u \bar u]^{\alpha_-^{\rm DB}} \, [1 + a_2^{\rm DB} C_2^{\alpha^{\rm DB}}(u - \bar u)]\,,
\end{equation}
$\alpha^{\rm DB} = 0.81$, $\alpha_-^{\rm DB} = \alpha^{\rm DB}-1/2$, $a_2^{\rm DB}=-0.12$; and
\begin{equation}
\label{phiKA}
\varphi_K^{\rm DB}(u;\zeta_2) = \,_m\phi_K^E(u)+ \,_m\phi_K^O(u)\,,
\end{equation}
with the functions defined in Eqs.\,\eqref{phimEO} and
\begin{equation}
\label{phiKB}
\begin{array}{lccccc}
\bar\alpha & \hat \alpha & a_2^{\bar\alpha} & a_1^{\hat\alpha} & a_3^{\hat\alpha} \\
1.42 & 1.14 & \phantom{-}0.074 & 0.076 & 0.011\\
\end{array}\,.
\end{equation}
These predictions agree with the best available estimates from numerical simulations of lQCD \cite{Shi:2014uwa,Segovia:2013eca,Cloet:2013tta}.

\begin{figure}[t]

\centerline{\includegraphics[width=0.90\linewidth]{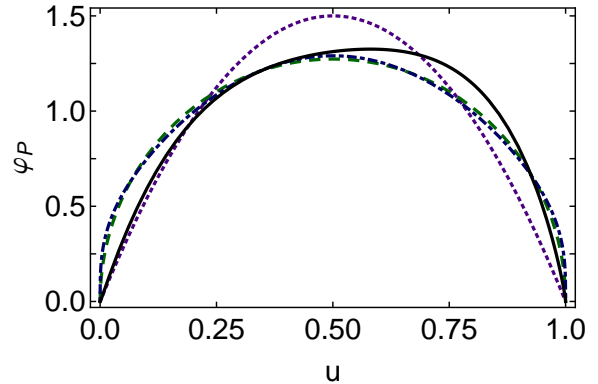}}

\caption{Two-particle twist-two PDAs, computed at $\zeta_2$: dot-dashed curve (dark blue) -- pion, Eq.\,\eqref{pionPDA2}; solid curve (black) -- kaon, Eqs.\,\eqref{phiKA}, \eqref{phiKB}; dashed curve (dark green) -- $\varphi_\pi^{\rm model}(u)$, Eq.\,\eqref{phiModel}; and dotted curve (indigo) -- asymptotic distribution, Eq.\,\eqref{phiasy}.
\label{T2piK}}
\end{figure}

We depict the curves of Eqs.\,\eqref{pionPDA2}--\eqref{phiKB} in Fig.\,\ref{T2piK}, and compare them with the asymptotic two-particle distribution, Eq.\,\eqref{phiasy}, and also with a model result \cite{Mikhailov:1986be,Brodsky:2006uqa}:
\begin{equation}
\label{phiModel}
\varphi_\pi^{\rm model}(u)=(8/\pi) \sqrt{u(1-u)}\,,
\end{equation}
which is practically indistinguishable from the DSE prediction.\footnote{\emph{N.B}.\ Whereas the scale that should be associated with the model analyses is poorly known, our result is computed at $\zeta_2$.  Therefore, the agreement emphasised here suggests that one is best advised to associate a scale of $\zeta_2$ with the model results, too.}   Notably, the DB prediction for the second moment of the pion's twist-two PDA is in agreement with earlier results from lQCD \cite{Braun:2006dg,Arthur:2010xf} and also confirmed by a more recent analysis \cite{Braun:2015axa}, \emph{viz}.
\begin{equation}
\langle u_\Delta^2 \rangle_{\varphi_\pi}^{\rm DB} = 0.25 \quad  {\rm cf.} \quad
\langle u_\Delta^2 \rangle_{\varphi_\pi}^{\rm lQCD} = 0.24 \pm 0.01\,.
\end{equation}

These results support observations made in the Introduction; namely, that at accessible energy scales, two-particle twist-two PDAs are broad, concave functions in which violations of $SU(3)$ flavour-symmetry are modulated by DCSB (since they are at the level of 14\%).  These issues are discussed further in Sec.\,\ref{SecLFCondensate}.

\subsection{Pseudoscalar, two-particle, twist-three}
\label{SecLFCondensate}
\subsubsection{Pion -- \mbox{$\mathbf \omega_\pi$}}
\begin{figure}[t]
\begin{minipage}[t]{0.5\textwidth}
\begin{minipage}{1.0\textwidth}
\centerline{\includegraphics[width=0.9\linewidth]{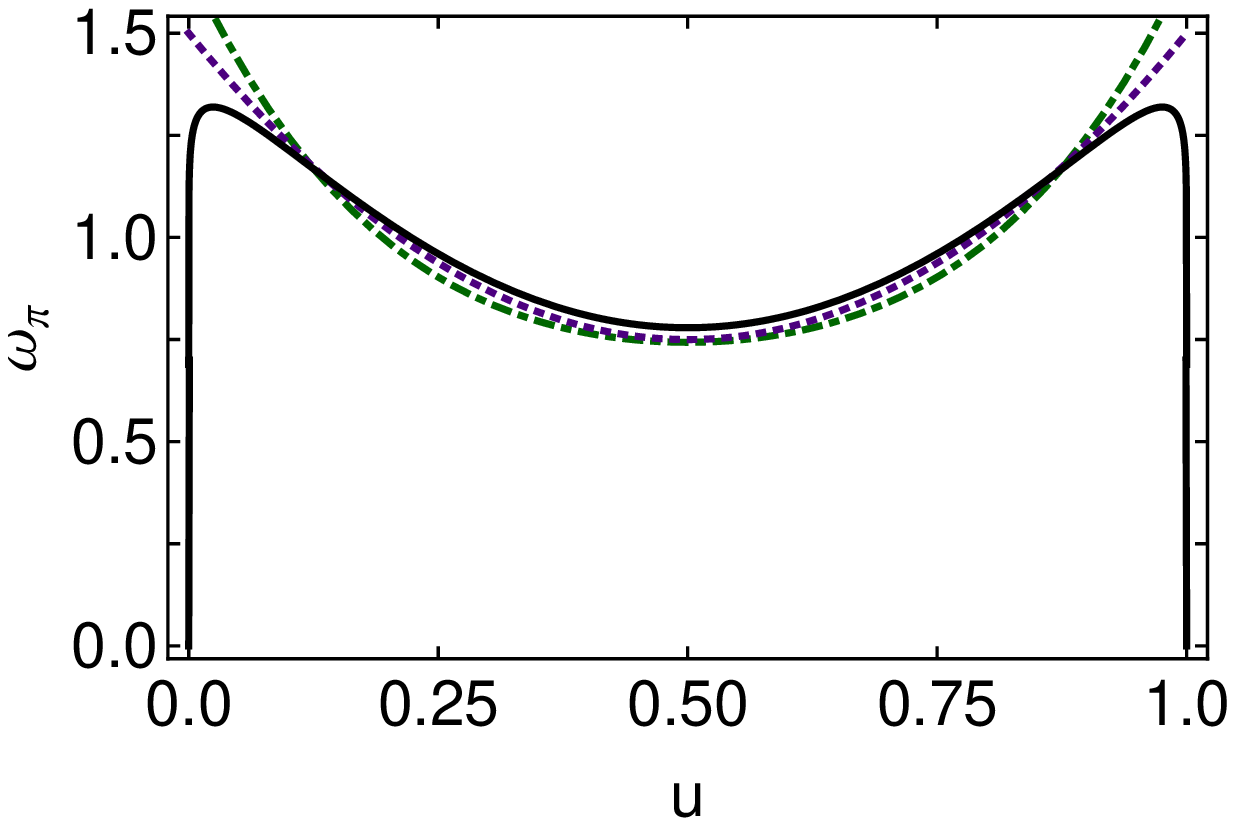}}
\end{minipage}
\begin{minipage}{1.0\textwidth}
\centerline{\includegraphics[width=0.9\linewidth]{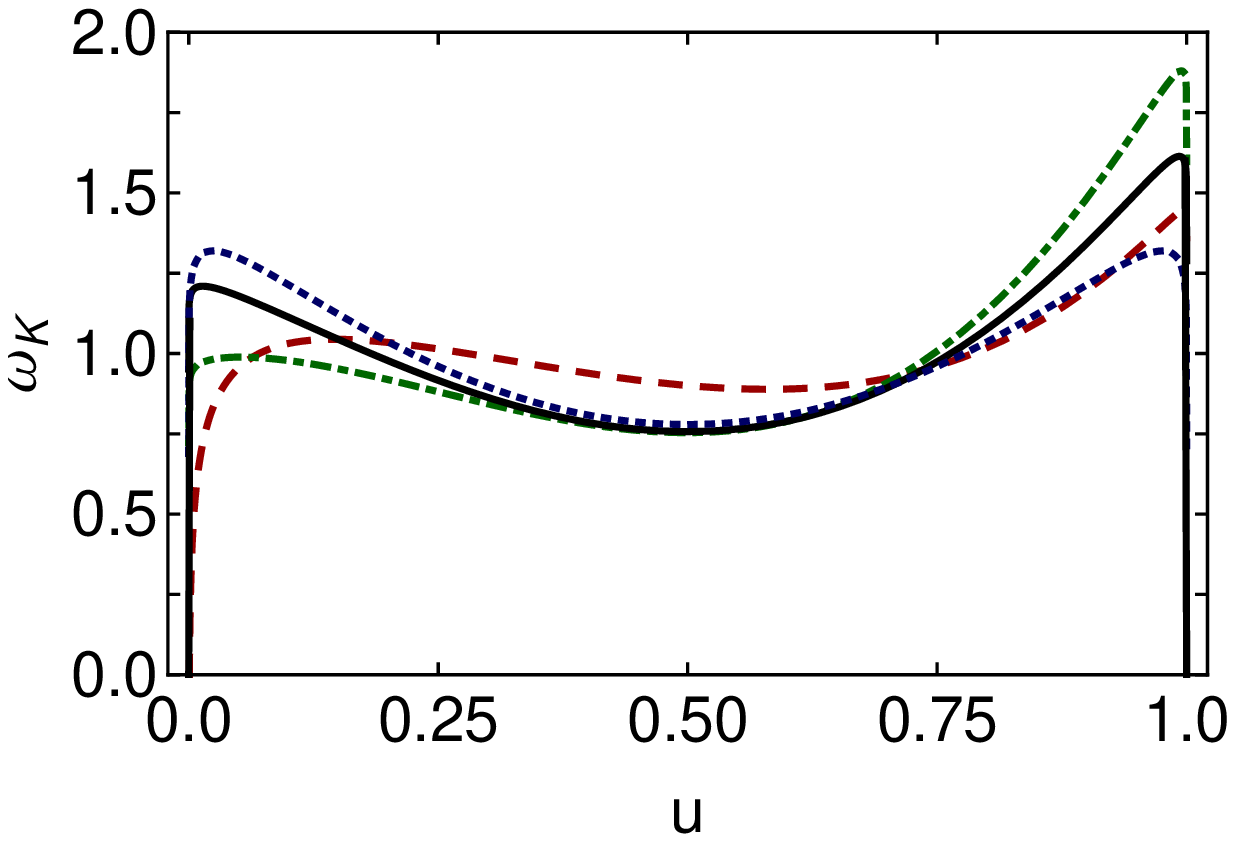}}
\end{minipage}
\end{minipage}
\caption{Pseudoscalar two-particle, twist-three PDAs, computed at $\zeta_2$.
\emph{Upper panel} -- pion: solid curve (black), prediction in Eq.\,\eqref{DSEomega}; dot-dashed curve (dark green), QCD sum rules estimate \cite{Ball:2006wn}; and
dotted curve (indigo), $\omega_P^{\rm asy}$ in Eq.\,\eqref{varphiasy}.
\emph{Lower panel} -- kaon:
solid curve (black), DB result;
dot-dashed curve (green), RL result;
dashed curve (red), QCD sum rules estimate \cite{Ball:2006wn};
and dotted curve (blue), pion prediction in Eq.\,\eqref{DSEomega}.
\label{T3PpiK}}
\end{figure}

The pion's two-particle twist-three PDA, $\omega_\pi(u)$, was computed and discussed in Ref.\,\cite{Chang:2013epa}.  In addition to playing an important role in the study of $B$-meson pionic decays \cite{Beneke:2003zv}, $\omega_\pi(u)$ may be viewed as describing the light-front distribution of the chiral condensate within the pion.  The chiral-limit prediction from Ref.\,\cite{Chang:2013epa} is depicted in the upper panel of Fig.\,\ref{T3PpiK}: the solid curve corresponds to
\begin{equation}
\label{DSEomega}
\omega_\pi(u;\zeta_2) =
N_\alpha [ u \bar u ]^{\alpha_-} [1+ a_2 C_2^{(\alpha)}(u - \bar u)]\,,
\end{equation}
with $\alpha = \nu-1/2$, $\nu=1.05$, $a_2=0.48$.  This is very close to $\omega_P^{\rm asy}(u)$ in Eq.\,\eqref{otherasy}, which is described by $\nu=1$, $a_2=1/2$.

It is important to be aware that only the $E_{P=\pi}(\ell;q)$ term in Eq.\,\eqref{BSK}  provides a nonzero contribution to the right-hand-side of Eq.\,\eqref{omegaE} when one removes the regularisation scale $\Lambda\to\infty$.  That is because $\lim_{\Lambda\to\infty} Z_4(\zeta,\Lambda) = 0$; and whilst the integral of the $E_{\pi}(\ell;q)$ term diverges with $\Lambda$ at precisely the rate required to produce a finite, nonzero, $\Lambda$-independent result, the terms $F_{\pi}(\ell;q)$, $G_{\pi}(\ell;q)$, $H_{\pi}(\ell;q)$ provide contributions to the integral that are finite as $\Lambda\to\infty$ and hence disappear when multiplied by a renormalisation constant which vanishes in this limit.

In addition, since the integral is dominated by the ultraviolet behaviour of the integrand, no difference in Bethe-Salpeter kernels at infrared momenta can have an impact.  Owing to the following quark-level Goldberger-Treiman relation \cite{Maris:1997hd,Qin:2014vya}:
\begin{equation}
f_\pi E_\pi(k;0) \stackrel{\hat m =0}{=} B_0(k^2)\,,
\label{gtlrelE}
\end{equation}
a pointwise expression of Goldstone's theorem in QCD, the chiral-limit prediction, Eq.\,\eqref{DSEomega}, is completely determined by the momentum-dependence of the scalar piece of the self-energy associated with the dressed-quark that is confined within the pion.  This momentum-dependence is the same in all DSE truncation schemes that preserve the one-loop renormalisation group properties of QCD, a fact that was confirmed in Ref.\,\cite{Chang:2013epa} by computing $\omega_\pi(x)$ in both RL and DB truncation and verifying that the results are identical.  These observations suggest strongly that the character of our result is a model-independent feature of strong-coupling QCD.

Notable, too, is the agreement evident in Fig.\,\ref{T3PpiK} between our prediction, Eq.\,\eqref{DSEomega}, and an earlier QCD sum rules result \cite{Ball:2006wn}.  (A quantitatively similar distribution is reported in Ref.\,\cite{Kopeliovich:2011rv}.) The discrepancy near the endpoints of the domain of support is understandable, given that just low-order moments can practically be constrained in a sum rules analysis and such moments possess little sensitivity to the behaviour of $\omega_\pi$ in the neighbourhood of the endpoints.  We judge that the generally good agreement with the prediction in Eq.\,\eqref{DSEomega} provides strong support for the model-independent nature of that result.  This is further emphasised by the fact that the estimate in Ref.\,\cite{Ball:2006wn} improves over an earlier calculation \cite{Ball:1998je} and, as gauged by the $L^1$-norm, the modern refinement shifts the earlier estimate toward the result in Eq.\,\eqref{DSEomega}.  Significantly, all results for $\omega_\pi(u)$ differ materially from $\omega_\pi^{\rm cl}(u)\equiv 1$.

It is appropriate at this point to remark that numerical results for $\omega_P(x)$ have recently been obtained using a light-front constituent-quark model \cite{Choi:2014ifm}.  Those results are in marked disagreement with all curves depicted in Fig.\,\ref{T3PpiK}: they have curvature of the opposite sign on almost the entire domain of support.  Consequently, the quark model results are in conflict with a model-independent prediction of QCD, which is deeply rooted in DCSB.  It is probable that this defect originates in the inability of constituent-quark models to veraciously express chiral symmetry and the pattern by which it is broken in QCD.  There are other, kindred examples, such as the failure of constituent-quark models \cite{Arndt:1999wx} to deliver zero as the chiral-limit value for the leptonic decay constant of excited-state pseudoscalar mesons, which is also a model-independent corollary of DCSB \cite{Volkov:1996br,Volkov:1999yi,Holl:2004fr,Holl:2005vu,Lucha:2006rq,%
McNeile:2006qy,Lucha:2007zz,Qin:2011xq,Rojas:2014aka,Ballon-Bayona:2014oma}.  This flaw is apparent in any approach that does not reliably incorporate and express the nature of chiral symmetry in QCD, \emph{e.g}.\ Ref.\,\cite{Mastropas:2014fsa}.

\subsubsection{Kaon -- \mbox{$\mathbf \omega_K$}}
We have computed the pseudoscalar two-particle, twist-three parton distribution amplitude for the negative-kaon using both the RL and DB truncations: for the reasons explained above, only the $E_{P=K}(\ell;q)$ term in Eq.\,\eqref{BSK} provides a nonzero contribution.  The results may be quoted in the form
\begin{equation}
\label{phiK5A}
\omega_K(u;\zeta_2) = \,_m\phi_K^E(u)+ \,_m\phi_K^O(u)\,,
\end{equation}
with the functions defined in Eqs.\,\eqref{phimEO} and
\begin{equation}
\label{phiK5B}
\begin{array}{l|cccccc}
& \bar\alpha & \hat \alpha & a_2^{\bar\alpha} & a_1^{\hat\alpha} & a_3^{\hat\alpha} \\\hline
\mbox{RL} & 0.52 & 0.51 & 0.49 & 0.29 & 0.19\\
\mbox{DB} & 0.52 & 0.52 & 0.48 & 0.13 & 0.08\\
\end{array}\,.
\end{equation}

\begin{table}[t]
\caption{Moments ($u_\Delta=2u-1$) of the $\pi$ and $K$-meson pseudoscalar two-particle, twist-three PDAs at $\zeta_2$, computed using Eqs.\,\eqref{DSEomega}, \eqref{phiK5A} and \eqref{phiK5B}.
We also list values obtained with $\omega = \omega_P^{\rm asy}$, Eq.\,\eqref{otherasy}, and computed using the QCD sum rules estimate for $\omega_K$ in Ref.\,\cite{Ball:2006wn}.
\emph{N.B}.\ For the kaon, $\omega_K^{\rm DB}$ is our most realistic result.
\label{momentsRLDB}
}
\begin{tabular*}
{\hsize}
{
l|
l@{\extracolsep{0ptplus1fil}}
l@{\extracolsep{0ptplus1fil}}
l@{\extracolsep{0ptplus1fil}}
l@{\extracolsep{0ptplus1fil}}
l@{\extracolsep{0ptplus1fil}}
l@{\extracolsep{0ptplus1fil}}}\hline
 $\langle u_\Delta^m \rangle$    & $m=1$ & $2$ & $3$ & $4$ & $5$ & $6$\\\hline
$\omega=\omega^{\rm asy}$ & 0 & 0.4 & 0 & 0.26 & 0 & 0.19 \\\hline
$\omega_\pi$  & 0 & 0.39 & 0 & 0.24 & 0 & 0.18 \\\hline
$\omega_K^{\rm RL}$  & 0.10 & 0.40 & 0.070 & 0.25 & 0.054 & 0.19 \\\hline
$\omega_K^{\rm DB}$  & 0.044 & 0.40 & 0.031 & 0.25 & 0.024 & 0.19 \\\hline
$\omega_K$\,\cite{Ball:2006wn}  & 0.041 & 0.36 & 0.038 & 0.22 & 0.033 & 0.16 \\\hline
\end{tabular*}
\end{table}

The predictions in Eqs.\,\eqref{phiK5A}, \eqref{phiK5B} are associated both with the moments listed in Table~\ref{momentsRLDB} and the solid- and dot-dashed-curves plotted in the lower panel of Fig.\,\ref{T3PpiK}: semi-quantitative agreement with a QCD sum rules estimate \cite{Ball:2006wn} is apparent.

There are a number of important messages that may be read from these results.  Plainly, as with the twist-two amplitude, the kaon distribution is skewed in favour of the heavier $s$-quark.  Here, however, since this amplitude describes the light-front distribution of the chiral condensate within the hadron \cite{Chang:2013epa}, the impact of the current-quark masses is most significant at the endpoints of the domain of support: the PDA is suppressed on $x<1/2$ and enhanced on $x>1/2$.  The distortion may be quantified by considering a ratio, \emph{viz}.\
\begin{equation}
\label{SU3omega}
\delta_{\omega_K} = \frac{\int_0^{\frac{1}{2}} d\bar u \, \omega_K(\bar u)}{\int_0^{\frac{1}{2}} du \, \omega_K(u)} =
\left\{
\begin{array}{ll}
1.28 & {\rm RL}\\
1.12 & {\rm DB}
\end{array}\right.\,,
\end{equation}
where $\bar u = 1-u$.  When the DB result for $\varphi_K(u)$ is employed, the analogous ratio evaluates to $\delta_{\varphi_K}^{\rm DB} = 1.14$.  It is thus evident that the magnitude of $SU(3)$ flavour-symmetry breaking in the kaon's pseudoscalar two-particle, twist-three distribution is similar to that in the twist-two PDA.

The magnitude of flavour symmetry-breaking exposed in Eq.\,\eqref{SU3omega} may also be compared with the 15\% shift in the peak of the kaon's valence $s$-quark parton distribution function, $s_v^K(x)$, relative to $u_v^K(x)$ \cite{Nguyen:2011jy} and the ratio of neutral- and charged-kaon electromagnetic form factors measured in $e^+e^-$ annihilation at $s_U=17.4\,$GeV$^2$ \cite{Seth:2013eaa}: $|F_{K_S K_L}(s_U)|/|F_{K_- K_+}(s_U)|\approx 0.12$.
By way of context, it is notable that the ratio of $s$-to-$u$ current-quark masses is approximately $27$ \cite{Beringer:1900zz}, whereas the ratio of nonperturbatively generated Euclidean constituent-quark masses is typically $1.5$ \cite{Chen:2012qr} and the ratio of leptonic decay constants $f_K/f_\pi \approx 1.2$ \cite{Beringer:1900zz}.  Both latter quantities are equivalent order parameters for DCSB.

Moreover, a DSE-based computation of leptonic decay constant ratios yields $f_{B_s}/f_B = 1.2$ \cite{Ivanov:2007cw}, in accord with a recent result from unquenched lattice-QCD $f_{B_s}/f_B=1.22(8)$ \cite{Christ:2014uea}, and the same DSE framework produces $f^+_{BK}(0)/f^+_{B\pi}(0)=1.21$ for the ratio of $B\to K,\pi$ semileptonic transition form factors at the maximum recoil point, a value that is typical for estimates of this quantity: the results in Refs.\,\cite{Melikhov:1997wp,Melikhov:2001zv,Faessler:2002ut,Ball:2004ye,%
Khodjamirian:2006st,Ebert:2006nz,Lu:2007sg} may be summarised as $f^+_{BK}(0)/f^+_{B\pi}(0)=1.26(5)$.

It is therefore apparent that, as with $\varphi_K(u)$, the flavour-dependence of DCSB rather than explicit chiral symmetry breaking is measured by the skewing of $\omega_K(u)$: $SU(3)$ flavour-symmetry breaking is far smaller than one might na\"ively have expected because DCSB impacts heavily on $u,d$- and $s$-quarks.

Looking closer at the results in the lower panel of Fig.\,\ref{T3PpiK}, one observes that the RL PDA is more skewed than the DB result, \emph{viz}.\ the RL truncation allocates a significantly larger fraction of the in-kaon condensate to its valence $s$-quark.  This feature is also highlighted by comparing the RL and DB results for the moments in Table~\ref{momentsRLDB}: the $m=1,3,5$ RL moments are noticeably larger than the odd moments obtained with the DB kernel.
This is readily understood.
RL-kernels ignore DCSB in the quark-gluon vertex.  Therefore, to describe a given body of phenomena, they must shift all DCSB strength into the infrared behaviour of the dressed-quark propagator, whilst nevertheless maintaining perturbative behaviour for $p^2>\zeta_2^2$.  This requires $M_{s,u}(p^2)$ to be unnaturally large at $p^2=0$ and then drop quickly with increasing $p^2$, behaviour which influences $\omega_K(u)$ via the Bethe-Salpeter equation.
In contrast, the DB-kernel builds DCSB into the quark-gluon vertex and its impact is therefore shared between more elements of a calculation.  Hence smaller values of $M_{s,u}(p^2=0)$ are capable of describing the same body of phenomena; and these dressed-masses need fall less rapidly in order to reach the asymptotic limits they share with the RL self-energies.  The DB kernel therefore produces a more balanced expression of DCSB's impact on a meson's Bethe-Salpeter wave function and hence the PDAs derived therefrom provide a more realistic expression of DCSB-induced skewing: DB provides the most realistic result \cite{Binosi:2014aea}.

The lower panel of Fig.\,\ref{T3PpiK} also facilitates a comparison between the kaon's pseudoscalar two-particle, twist-three PDA and that obtained for the pion using the same kernel \cite{Chang:2013pq}.  Setting the asymmetry of the kaon's distribution aside, the qualitative character of the distributions is the same: they are both maximal at the endpoints of the domain of support.  This is a definitive signature of the Goldstone boson structure of these mesons, which shows that the chiral condensate is primarily located in components of the pseudoscalar meson wave functions that express correlations with large relative momenta, a feature which ensures, \emph{inter alia},  that light-front longitudinal zero modes do not play a material role in forming the chiral condensate \cite{Chang:2013epa}.

\subsection{Pseudotensor, two-particle twist-three}
\label{SecPT}
\begin{figure}[t]

\centerline{\includegraphics[width=0.90\linewidth]{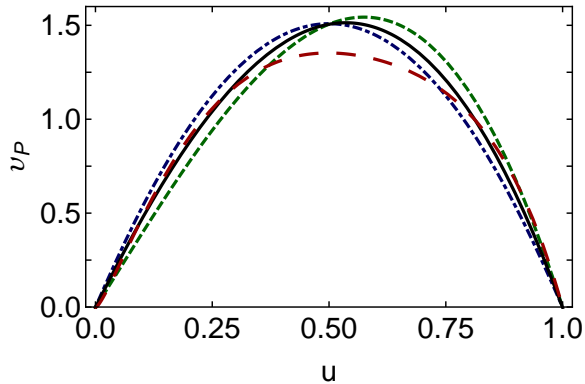}}

\caption{Pseudotensor two-particle, twist-three PDAs, computed at $\zeta_2$:
dot-dashed curve (dark blue) -- pion, $\upsilon_\pi(u)$ in Eq.\,\eqref{pionPDA2};
solid curve (black) -- kaon in DB truncation, $\upsilon_K^{\rm DB}(u)$ in Eqs.\,\eqref{phiKA}, \eqref{phiKB};
dashed curve (dark green) -- kaon in RL truncation $\upsilon_K^{\rm RL}(u)$ in Eqs.\,\eqref{phiKA}, \eqref{phiKB};
and
long-dashed curve (red) -- QCD sum rules result for the kaon from Ref.\,\cite{Ball:2006wn}.
We do not plot the asymptotic form in Eq.\,\eqref{upsilonasy} because it is effectively indistinguishable from our prediction for the pion.
\label{T3PTpiK}}
\end{figure}

We have also computed the distribution $\upsilon_P(u)$ defined via Eq.\,\eqref{upsilonE}, using the same technique employed for the other PDAs.  As was the case for the pseudoscalar distribution amplitudes, the value of the integral in Eq.\,\eqref{upsilonEE} is dominated by the ultraviolet behaviour of $E_{P}(\ell;q)$: the other elements in the pseudoscalar meson Bethe-Salpeter amplitude, Eq.\,\eqref{BSK}, play no role.  Once more, therefore, there is no model dependence in the pion computation and the RL and DB results are identical:
\begin{equation}
\label{DSEpiUpsilon}
\upsilon_\pi(u;\zeta_2) = 6 u\bar u [1-0.0048 \, C_2^{(3/2)}(2u-1)]\,,
\end{equation}
which is not meaningfully distinguishable from $\upsilon_\pi^{\rm asy}(u)$ in Eq.\,\eqref{upsilonasy}.  This function is depicted in Fig.\,\ref{T3PTpiK}.

Our predictions for the kaon are also simple.  Their pointwise forms are well described by:
\begin{equation}
\label{DSEKUpsilonA}
\upsilon_K^{\rm T}(u) = 6 u\bar u\,[1+\sum_{i=1}^2 a_i^{\rm T} C_i^{(3/2)}(2u-1)]\,,
\end{equation}
where T=RL, DB and
\begin{equation}
\label{DSEKUpsilonB}
\begin{array}{lll}
    & a_1 & \phantom{-}a_2 \\
{\rm RL}  & 0.11 & -0.0035 \\
{\rm DB}  & 0.049 & -0.0034
\end{array}\,.
\end{equation}
These functions are depicted in Fig.\,\ref{T3PTpiK} and are associated with the moments in Table~\ref{momentsUpsilon}.

In this instance, the distortion of the distributions can readily be measured by the shift in peak location relative to that of the pion: the RL amplitude peaks at $u=0.57$ and DB at $u=0.54$.  One may also use $\delta_{\upsilon_K}$, defined via obvious analogy with Eq.\,\eqref{SU3omega}:
\begin{equation}
\delta_{\upsilon_K}^{\rm RL}=1.28\,, \quad \delta_{\upsilon_K}^{\rm DB}=1.12\,.
\end{equation}
Thus, in the more realistic (DB) case, the breaking of $SU(3)$ flavour-symmetry in $\upsilon_K$ is 12\%, as it is within the other kaon PDAs calculated herein.  Hence, unsurprisingly, the effect is once again modulated by the current-quark mass dependence of DCSB.

In Fig.\,\ref{T3PTpiK} we also display a QCD sum rules estimate for $\upsilon_K$ \cite{Ball:2006wn}.  In this case, too, the result obtained via that route is semi-quantitatively in agreement with our prediction, possessing a similar level of distortion toward $u=1$: 9\% as measured by $\delta_{\upsilon_K}$.  (The sum rules result for the pion distribution is symmetric around $u=1/2$ but otherwise qualitatively similar in shape and magnitude to the sum-rules kaon distribution, so it is not drawn.)

\section{Conclusion}
\label{epilogue}
We described calculations of the pointwise form for all kaon and pion dressed-quark (two-particle) parton distribution amplitudes (PDAs) to twist-three.  These computations have become possible owing to the development and use of novel methods for the algebraic interpolation of dressed-quark propagators and meson Bethe-Salpeter amplitudes, which are based on the notion of generalised perturbation theory integral representations \cite{Chang:2013pq,Nakanishi:1971}.

Before providing details, we list here our three primary conclusions.  Namely, only the pseudotensor PDA can reasonably be approximated by its conformal limit.  At any realistic energy scale, the twist-two and pseudoscalar twist-three PDAs differ markedly from the functions which are commonly associated with their forms in QCD's conformal limit.  Moreover, in all amplitudes studied, $SU(3)$ flavour-symmetry breaking is typically a 13\% effect, the scale of which is determined by nonperturbative dynamics, \emph{viz}.\ the current-quark-mass dependence of dynamical chiral symmetry breaking (DCSB).  The heavier-quark is always favoured by this distortion, \emph{e.g}.\ support is shifted toward $u=1$ in the $K^-$.  It follows that kaon and pion PDAs with the properties elucidated herein should serve as the basis for future attempts to access CP violation in the Standard Model.

\begin{table}[t]
\caption{Moments ($u_\Delta=2u-1$) of the $\pi$ and $K$-meson pseudotensor two-particle, twist-three PDAs at $\zeta_2$, computed using Eqs.\,\eqref{DSEpiUpsilon}, \eqref{DSEKUpsilonA} and \eqref{DSEKUpsilonB}.
We also list values obtained with $\upsilon_P = \upsilon_P^{\rm asy}$, Eq.\,\eqref{upsilonasy}, and computed using the QCD sum rules estimate for $\upsilon_K$ in Ref.\,\cite{Ball:2006wn}.
\emph{N.B}.\ For the kaon, $\upsilon_K^{\rm DB}$ is our most realistic result.
\label{momentsUpsilon}
}
\begin{tabular*}
{\hsize}
{
l|
l@{\extracolsep{0ptplus1fil}}
l@{\extracolsep{0ptplus1fil}}
l@{\extracolsep{0ptplus1fil}}
l@{\extracolsep{0ptplus1fil}}
l@{\extracolsep{0ptplus1fil}}
l@{\extracolsep{0ptplus1fil}}}\hline
 $\langle u_\Delta^m \rangle$    & $m=1$ & $2$ & $3$ & $4$ & $5$ & $6$\\\hline
$\upsilon=\upsilon^{\rm asy}$ & 0 & 0.2 & 0 & 0.086 & 0 & 0.048 \\\hline
$\upsilon_\pi$  & 0 & 0.20 & 0 & 0.085 & 0 & 0.047 \\\hline
$\upsilon_K^{\rm RL}$  & 0.065 & 0.20 & 0.028 & 0.085 & 0.015 & 0.047 \\\hline
$\upsilon_K^{\rm DB}$  & 0.029 & 0.20 & 0.013 & 0.085 & 0.0070 & 0.047 \\\hline
$\upsilon_K$\,\cite{Ball:2006wn}  & 0.030 & 0.20 & 0.017 & 0.088 & 0.011 & 0.049 \\\hline
\end{tabular*}
\end{table}

Turning now to specifics, the kaon and pion twist-two PDAs are broad, concave functions, in which the violation of $SU(3)$ flavour-symmetry is a 12-16\% effect when measured by the appearance of asymmetry in the kaon's PDA \cite{Chang:2013pq,Cloet:2013tta,Chang:2013nia,Segovia:2013eca,Shi:2014uwa}.  All features of these PDAs are modulated by DCSB and they cannot be approximated satisfactorily by  $\varphi^{\rm cl}_P(u)= 6 u(1-u)$, the form associated with QCD's conformal limit, at any energy scale achievable with terrestrial facilities.

The kaon and pion dressed-quark pseudoscalar twist-three PDAs, $\omega_P(u)$, are particularly interesting.  Algebraic analyses based on simple input [Eqs.\,\eqref{NakanishiASY}] produce asymptotic forms for the twist-two and pseudotensor twist-three PDAs that both coincide with the expressions anticipated of QCD's conformal limit, $\varphi^{\rm cl}(u)=6 u (1-u)$ [Eqs.\,\eqref{upsilonasy}, \eqref{varphiasy}].  However, kindred analysis for the pseudoscalar twist-three PDA produces $\omega_P^{\rm asy}(u)=1+(1/2)C_2^{(1/2)}(2u-1)$ [Eq.\,\eqref{otherasy}].  This form locates significant strength at the endpoints of the distribution's domain of support, depleting the central region, in marked contrast to the function which is associated with this PDA in QCD's conformal limit: $\omega_P^{\rm cl}(u)\equiv 1$.  Notwithstanding this, our numerical results for $\omega_{P=\pi,K}^{\rm asy}(u)$ are best understood when referred to $\omega_P^{\rm asy}(u)$ as the benchmark [Fig.\,\ref{T3PpiK}].  The same is true of the amplitude estimated when conformal invariance is used as the guiding principle and QCD sum rules are employed to estimate the relevant mass-scale parameters.  Thus, in this case, too, if $\omega_P^{\rm cl}(u)$ is truly the conformal-limit result in QCD, then it is irrelevant to contemporary and foreseeable experiments.  Furthermore, as with the kaon's twist-two PDA, $SU(3)$ flavour-symmetry breaking in the dressed-quark pseudoscalar, twist-three PDA, $\omega_P(u)$, is a 12\% effect when measured by the mass-induced distortion of the PDA.

The pseudotensor dressed-quark twist-three PDAs are the simplest of the quantities we considered [Fig.\,\ref{T3PTpiK}].  The computed pion result is almost identical to the functional form associated with the conformal-limit, $\varphi^{\rm cl}_P(u)$; and the kaon's PDA is a modestly asymmetrised version of $\varphi_P^{\rm cl}(u)$, with violation of $SU(3)$ flavour-symmetry again at the level of 12\%.

It is worth reiterating some of the advantages in using the Dyson-Schwinger Equation (DSE) approach in studies such as this.  For example, the framework preserves the one-loop renormalisation group behaviour of QCD, so that current-quark masses have a direct connection with the parameters in QCD's action and the dressed-quark mass-functions, $M_{s,u}(p^2)$, are independent of the renormalisation point.  Unlike other approaches to nonperturbative phenomena in continuum QCD, the renormalisation point can be fixed unambiguously, as in lattice-QCD: it is not a parameter to be identified with some poorly determined ``typical hadronic scale.''  Moreover, one is not restricted to estimating a few low-order moments of the PDA.  In working in the continuum and computing Bethe-Salpeter wave functions directly, the DSEs enable one to deliver predictions for the pointwise behaviour of PDAs on the full domain $u\in [0,1]$.  Importantly, those predictions are parameter-free and unify a meson's PDAs with a diverse range of apparently distinct phenomena.

A coherent picture has now emerged.  Modern DSE studies predict PDAs for light-quark mesons that are typically quite different from their conformal limits and these differences are a clean expression of DCSB on the light front.  Notably, where a comparison is possible, the DSE results are consistent with those determined via contemporary numerical simulations of lattice-regularised QCD.  A new paradigm thus presents itself, from which it follows that at energy scales accessible with existing and foreseeable facilities, one may arrive at reliable expectations for the outcome of experiments by using these ``strongly dressed'' PDAs in formulae for hard exclusive processes.  Following this procedure, any discrepancies will be significantly smaller than those produced by using the conformal-limit PDAs in such formulae.  Moreover, the magnitude of any disagreement will either be a better estimate of higher-order, higher-twist effects or provide more realistic constraints on the Standard Model.

\begin{acknowledgments}
We thank I.\,C.~Clo\"et, S.-X.~Qin, J.~Segovia Gonzalez, P.\,C.~Tandy, A.\,W.~Thomas and S.-L.~Wan for insightful comments.
Work supported by:
the National Natural Science Foundation of China (grant nos.\ 11275097
and 11475085); the National Basic Research Programme of China (grant no.\ 2012CB921504);
the Fundamental Research Funds for the Central Universities Programme of China, (grant no.\ WK2030040050);
University of Adelaide and Australian Research Council through grant no.~FL0992247;
U.S.\ Department of Energy, Office of Science, Office of Nuclear Physics, under contract no.~DE-AC02-06CH11357;
and For\-schungs\-zentrum J\"ulich GmbH.
\end{acknowledgments}

\appendix
\setcounter{figure}{0}
\setcounter{table}{0}
\renewcommand{\thefigure}{\Alph{section}.\arabic{figure}}
\renewcommand{\thetable}{\Alph{section}.\arabic{table}}

\section{Interpolations of propagators and Bethe-Salpeter amplitudes}
\label{sec:interpolations}
Here we describe the interpolations used in our evaluation of the moments in Eq.\,\eqref{momentsE}.  There are two sets of results to consider; viz., those obtained in RL truncation and those produced by the DB truncation.  The interaction in Ref.\,\cite{Qin:2011dd} has one parameter $m_g^3 := D\omega$ because with $m_g = \,$constant, light-quark observables are independent of the value of $\omega \in [0.4,0.6]\,$GeV.  We use $\omega =0.5\,$GeV.

In RL truncation, with $m_g = 0.82\,$GeV and renormalisation point invariant current-quark masses $\hat m_{u/d}=6.8\,$MeV, $\hat m_s = 162\,$MeV, which correspond to the following one-loop evolved masses $m_{u/d}^{\zeta=2\,{\rm GeV}} = 4.7\,$MeV, $m_{s}^{\zeta=2\,{\rm GeV}} =112\,$MeV, we obtain $m_\pi=0.14\,$GeV, $f_\pi=0.093\,$GeV and $m_K=0.49$GeV, $f_K=0.11\,$GeV.

Using the DB truncation with $m_g = 0.55
\,$GeV, we obtain $m_\pi=0.14\,$GeV, $m_K=0.50$GeV from renormalisation point invariant current-quark masses $\hat m_{u/d}=4.4\,$MeV, $\hat m_s = 90\,$MeV, which yield $m_{u/d}^{\zeta=2\,{\rm GeV}} = 3.0\,$MeV, $m_{s}^{\zeta=2\,{\rm GeV}} =62\,$MeV and produce the following values of the dressed-quark mass $M_u(\zeta_2)=4.3\,$MeV, $M_s(\zeta_2)=89\,$MeV, which are in fair agreement with modern lattice estimates \cite{Carrasco:2014cwa}.

\begin{table}[t]
\caption{Representation parameters. Eq.\,\protect\eqref{Spfit} -- the pair $(x,y)$ represents the complex number $x+ i y$.  (Dimensioned quantities in GeV).
\label{paramsquark}
}
\begin{center}
\begin{tabular*}
{\hsize}
{
l|@{\extracolsep{0ptplus1fil}}
c@{\extracolsep{0ptplus1fil}}
c@{\extracolsep{0ptplus1fil}}
c@{\extracolsep{0ptplus1fil}}
c@{\extracolsep{0ptplus1fil}}
c@{\extracolsep{0ptplus1fil}}}\hline
 RL & $z_1$ & $m_1$  & $z_s$ & $m_2$ \\

 $u$& $(0.38,0.71)$ & $(0.71,0.22)$ & $(0.14,0)$ & $(-0.78,0.75)$ \\
 $s$& $(0.42,0.32)$ & $(0.80,0.41)$ & $(0.12,0)$ & $(-1.26,0.63)$ \\
 DB & $z_1$ & $m_1$  & $z_s$ & $m_2$ \\
 $u$& $(0.42,0.24)$ & $(0.44,0.19)$ & $(0.13,0.07)$ & $(-0.76,0.60)$ \\
 $s$& $(0.43,0.30)$ & $(0.55,0.22)$ & $(0.12,0.11)$ & $(-0.83,0.42)$ \\\hline
\end{tabular*}
\end{center}
\end{table}

In interpolating the results from either truncation, the dressed-quark propagators are represented as \cite{Bhagwat:2002tx}
\begin{equation}
S_f(p) = \sum_{j=1}^{j_m}\bigg[ \frac{z_j^f}{i \gamma\cdot p + m_j^f}+\frac{z_j^{f\ast}}{i \gamma \cdot p + m_j^{f\ast}}\bigg], \label{Spfit}
\end{equation}
with $\Im m_j \neq 0$ $\forall j$, so that $\sigma_{V,S}$ are meromorphic functions with no poles on the real $p^2$-axis, a feature consistent with confinement \cite{Bashir:2012fs}.  We find that $j_m=2$ is adequate; and the interpolation parameters are listed in Table~\ref{paramsquark}.  (Compared with Ref.\,\cite{Shi:2014uwa}, we have made a minor modification of the RL $s$-quark parameters, which slightly improves the quality of the fit but otherwise has no observable impact.)

The Bethe-Salpeter amplitude for a pseudoscalar meson is given in Eq.\,\eqref{BSK}.  For the pion, ${\mathpzc F}_1(\ell;q)\equiv 0$ in Eq.\,\eqref{decom} and it is natural to choose $\eta=1/2$.  In this case we represent the scalar functions in Eq.\,\eqref{BSK} $({\cal F}=E,F,G)$ by
{\allowdisplaybreaks
\begin{eqnarray}
{\cal F}(\ell;q) &=& {\cal F}^{\rm i}(\ell;q) + {\cal F}^{\rm u}(\ell;q) \,, \\
\nonumber {\cal F}^{\rm i}(\ell;q) & = & c_{\cal F}^{\rm i}\int_{-1}^1 \! dz \, \varsigma_{\nu^{\rm i}_{\cal F}}(z) \bigg[
a_{\cal F} \hat\Delta_{\Lambda^{\rm i}_{{\cal F}}}^4(\ell_z^2) \\
&& \rule{7em}{0ex}
+ a^-_{\cal F} \hat\Delta_{\Lambda^{\rm i}_{\cal F}}^5(\ell_z^2)
\bigg], \label{Fifit}\\
E^{\rm u}(\ell;q) & = & c_{E}^{\rm u} \int_{-1}^1 \! dz \, \varsigma_{\nu^{\rm u}_E}(z)\,
 \hat \Delta_{\Lambda^{\rm u}_{E}}(\ell_z^2)\,,\\
F^{\rm u}(\ell;q) & = & c_{F}^{\rm u} \int_{-1}^1 \! dz \, \varsigma_{\nu^{\rm u}_F}(z)\,
 \Lambda_F^{\rm u} k^2 \Delta_{\Lambda^{\rm u}_{F}}^2(\ell_z^2)\,,\\
G^{\rm u}(\ell;q) & = & c_{G}^{\rm u} \int_{-1}^1 \! dz \, \varsigma_{\nu^{\rm u}_G}(z)\,
 \Lambda_G^{\rm u}\Delta_{\Lambda^{\rm u}_{G}}^2(\ell_z^2)\,, \label{Gufit}
\end{eqnarray}}
\hspace*{-0.5\parindent}with $\hat \Delta_\Lambda(s) = \Lambda^2 \Delta_\Lambda(s)$, $\ell_z^2=\ell^2+z \ell\cdot q$, $a^-_E = 1 - a_E$, $a^-_F = 1/\Lambda_F^{\rm i} - a_F$, $a^-_G = 1/[\Lambda_G^{\rm i}]^3 - a_G$.  $H(\ell;q)$ is small, has little impact, and is thus neglected.  The interpolation parameters that fit our numerical results for the Bethe-Salpeter amplitudes are given in Table~\ref{Table:parameters}.  They were obtained elsewhere \cite{Chang:2013pq} through a least-squares fit to the Chebyshev moments
\begin{equation}
{\cal F}_n(\ell^2) = \frac{2}{\pi}\int_{-1}^{1}\!dx\, \sqrt{1-x^2} {\cal F}(\ell;q) U_n(x)\,,
\end{equation}
with $n=0,2$, where $U_n(x)$ is an order-$n$ Chebyshev polynomial of the second kind. Owing to $O(4)$ invariance, one may define $x = \hat k\cdot q/iQ$, with $\hat k^2=1$ and $q=(0,0,Q,i Q)$.  Only results obtained using the DB kernel for the pion are truly relevant herein and we therefore only list interpolation parameters for that case.  (\emph{N.B}.\ The overall multiplicative factor resulting from canonical normalisation of $\Gamma_\pi$ is not included in Table~\ref{Table:parameters}.)

\begin{table}[t]
\caption{Representation parameters. Eqs.\,\protect\eqref{Fifit}--\protect\eqref{Gufit}.  (Dimensioned quantities in GeV).
\label{Table:parameters}
}
\begin{center}
%

\begin{tabular*}
{\hsize}
{
l@{\extracolsep{0ptplus1fil}}
l@{\extracolsep{0ptplus1fil}}
c@{\extracolsep{0ptplus1fil}}
c@{\extracolsep{0ptplus1fil}}
c@{\extracolsep{0ptplus1fil}}
c@{\extracolsep{0ptplus1fil}}
c@{\extracolsep{0ptplus1fil}}
c@{\extracolsep{0ptplus1fil}}
c@{\extracolsep{0ptplus1fil}}
c@{\extracolsep{0ptplus1fil}}}\hline
    & & $c^{\rm i}$ & $c^{u}$ & $\phantom{-}\nu^{\rm i}$ & $\nu^{\rm u}$ & $a$\phantom{00} & $\Lambda^{\rm i}$ & $\Lambda^{\rm u}$\\\hline
DB: & E & $1 - c^{u}_E$ & $0.08$ & $-0.70$ & 1.08
    & 3.0\phantom{$/[\Lambda^{\rm i}_G]^3$} & 1.41 & 1.0\\
    & F & \phantom{-}0.55 & $c^{\rm u}_E/10$ & $\phantom{-}0.40$ & 0.0
    & 3.0$/\Lambda^{\rm i}_{F}$\phantom{00} & 1.13 & 1.0 \\
& G & $-0.094$ & 2$\,c^{\rm u}_F$ & $\phantom{-}\nu^{\rm i}_F$ & 0.0 & 1.0$/[\Lambda^{\rm i}_G]^3$ & 0.79 & 1.0 \\\hline
\end{tabular*}
\end{center}

\vspace*{-4ex}

\end{table}

\begin{table}[t]
\caption{Representation parameters associated with Eqs.\,\eqref{BSK}, \eqref{decom}, \eqref{fit}. (Dimensioned quantities in GeV.  Omitted quantities are zero or unused.)
\label{BSAparameters}
}
\begin{center}
\begin{tabular*}
{\hsize}
{
r|@{\extracolsep{0ptplus1fil}}
c@{\extracolsep{0ptplus1fil}}
c@{\extracolsep{0ptplus1fil}}
c@{\extracolsep{0ptplus1fil}}
c@{\extracolsep{0ptplus1fil}}
c@{\extracolsep{0ptplus1fil}}
c@{\extracolsep{0ptplus1fil}}}\hline
 RL & $E_0$ & $E_1$  & $F_0$ & $F_1$ & $G_0$ & $G_1$ \\\hline
 $\nu_{0}$& $-0.71$ & $0.17$ & $1.33$ & 5.62 & $1.0$ & -0.1 \\
 $\nu_1$  &         &        &        &      & $-0.7$&      \\
 $\nu_{2}$& $1.0$ & $2.0$ & $0.0$ & $0.0$ & $0.0$ & $0.0$\\
 $U_0$ & $1.0$ & $0.7$& $0.42$  & 0.21 & 0.0 & $0.28$ \\
 $U_1$ &       &      &         &      & 0.25&        \\
 $10^3 U_2$& 6.83 & 0.36 & 0.90  & 0.01 & -0.01 & 0.70 \\
 $n_0$& $5$ & 8& $5$ & 8 & 10 & 6 \\
 $n_1$&     &  &     &   & 12 &   \\
 $n_2$& $1$ &$ 2$& $1$ &$ 2$ &$ 2$ &$ 2$\\
 $\Lambda$& $1.8$ & $2.0$& $1.5$ & 1.6 & 2.1 & $1.5$ \\\hline
 DB & $E_0$ & $E_1$  & $F_0$ & $F_1$ & $G_0$ & $G_1$ \\\hline
$\nu_{0}$& $-0.54$ & $-0.1$ & $-0.01$ & 1.6 & 1.5 & $3.0$\\
$\nu_{1}$& $-0.7$ & $-0.4$& $-0.7$ & 0.8 &   & 3.0 \\
$\nu_{2}$& $1.0$ & $2.0$ & $0.0$ & $0.0$ & $0.0$ & $0.0$\\
$U_0$ & $1.0$ & $0.22$& $0.56$ & 0.11 & -0.058 & $0.12$ \\
$U_1$ & -2.0 & -0.5& -0.3  & -0.65 &    & -1.5\\
$10^2U_2$& 2.5 & 0.052 & 0.39& 0.001 & 0.049 & -0.60 \\
$n_0$& $4$ & 8& $4$ & 10 & 5 & 8 \\
$n_1$& $5$ & $12$& $6$ & 12 &   & 10 \\
$n_2$& $1$ &$ 2$& $1$ &$ 2$ & 2 & 2\\
$\Lambda$& $1.35$ & $1.7$& $1.2$ & 1.45 & 0.8 & $1.1$ \\\hline
\end{tabular*}
\end{center}
\end{table}

Here it is worth noting another detail associated with the generalised spectral representations.  DSE kernels that preserve the one-loop renormalisation group behaviour of QCD will necessarily generate propagators and Bethe-Salpeter amplitudes with a nonzero anomalous dimension $\gamma_F$, where $F$ labels the object concerned.  Consequently, the spectral representation must be capable of describing functions of $\mathpzc{s}=p^2/\Lambda_{\rm QCD}^2$ that exhibit $\ln^{-\gamma_F}[\mathpzc{s}]$ behaviour for $\mathpzc{s}\gg 1$.  This is readily achieved by noting that
\begin{equation}
\label{logfactor}
\ln^{-\gamma_F} [D(\mathpzc{s})]
= \frac{1}{\Gamma(\gamma_F)} \int_0^\infty \! dz\, z^{\gamma_F-1}
\frac{1}{[D(\mathpzc{s})]^z}\,,
\end{equation}
where $D(\mathpzc{s})$ is some function.  Such a factor can be multiplied into any existing spectral representation in order to achieve the required ultraviolet behaviour.

In connection with the dressed-quark two-particle twist-three distributions considered herein, it is the anomalous dimension of the dressed-quark mass-function that must properly be represented: $\gamma_F\to\gamma_m=12/25$ in the RL and DB kernels.  Owing to Eq.\,\eqref{gtlrelE}, this also affects the pion and kaon Bethe-Salpeter amplitudes.  In the integrands describing $E^{\rm u}(\ell;q)$ above, we therefore include a multiplicative factor
\begin{equation}
\label{factorln}
\big(\ln\Lambda_{\rm u}^2/\ln [\ell^2+\Lambda_{\rm u}^2]\big)^{1-\gamma_m}\,.
\end{equation}

An analogous procedure is followed for the kaon's Bethe-Salpeter amplitude.  In this instance, ${\mathpzc F}_1(\ell;q)\neq 0$ in Eq.\,\eqref{decom}.  Hence, interpolations must be found for both ${\mathpzc F}_{0,1}(\ell;q)$.  The following forms are flexible enough to allow a satisfactory representation of the numerical solutions to the Bethe-Salpeter equations:
{\allowdisplaybreaks
\begin{align}
\nonumber \mathcal{F}_j(\ell;q)&=
\int_{-1}^{1}d\alpha \, \varsigma_0(\alpha) \frac{(U_0 -U_1-U_2)\Lambda_j^{2 n_0}}{(\ell^2+\alpha\, \ell \cdot q+\Lambda_j^2)^{n_0}} \nonumber \\
\nonumber
& +\int_{-1}^{1}d\alpha \, \varsigma_1(\alpha) \frac{U_1\Lambda_j^{2 n_1}}{(\ell^2+\alpha\, \ell \cdot q+\Lambda_j^2)^{n_1}}
\\
&+\int_{-1}^{1}d\alpha \, \varsigma_2(\alpha) \frac{U_2\Lambda_j^{2 n_2}}{(\ell^2+\alpha\, \ell \cdot q+\Lambda_j^2)^{n_2}}\,,
\label{fit}
\end{align}}
\hspace*{-0.5\parindent}where $\varsigma_{\nu_i}(\alpha)$ is obtained via Eq.\,\eqref{eq:sim3}.  For the reasons described above, the multiplicative factor of Eq.\,\eqref{factorln} is included in the $E_{0,1}$-integrands containing $\varsigma_2(\alpha)$.  Interpolation parameters for each function ${\mathpzc F}_{0,1}(\ell;q)$ are determined via a least-squares fit to that function's $n=0,2$ Chebyshev moments.  The resulting parameter values are listed in Table~\ref{BSAparameters}.  (\emph{N.B}.\ The overall multiplicative factor resulting from canonical normalisation of $\Gamma_K$ is omitted.  The function $H_K$ is, too, because it does not have a noticeable effect on our results.)


\end{document}